\documentclass[journal]{IEEEtran}
\ifCLASSINFOpdf
\else
\fi

\usepackage{amsmath}
\usepackage{amsfonts}
\usepackage{cases}
\usepackage{amssymb}
\usepackage{color}
\usepackage{graphicx}
\newtheorem{lemma}{Lemma}
\newtheorem{remark}{Remark}
\newtheorem{assump}{Assumption}
\newtheorem{Def}{Definition}
\newtheorem{theorem}{Theorem}{}
\usepackage{cite}
\usepackage{lineno} % 这个包就是latex自带的添加行号的包了

\begin{document}
%\pagewiselinenumbers% 按页重新编号 
%\switchlinenumbers	% 双栏

	%
	% paper title
	% Titles are generally capitalized except for words such as a, an, and, as,
	% at, but, by, for, in, nor, of, on, or, the, to and up, which are usually
	% not capitalized unless they are the first or last word of the title.
	% Linebreaks \\ can be used within to get better formatting as desired.
	% Do not put math or special symbols in the title.
	\title{Optimal control of nonlinear systems with unsymmetrical input constraints and its application to the  UAV circumnavigation problem}
	%
	%
	% author names and IEEE memberships
	% note positions of commas and nonbreaking spaces ( ~ ) LaTeX will not break
	% a structure at a ~ so this keeps an author's name from being broken across
	% two lines.
	% use \thanks{} to gain access to the first footnote area
	% a separate \thanks must be used for each paragraph as LaTeX2e's \thanks
	% was not built to handle multiple paragraphs
	%

	\author{Yangguang~Yu,
		Xiangke~Wang,~\IEEEmembership{Senior~Member,~IEEE,}
		Zhiyong~Sun,~\IEEEmembership{Member,~IEEE,}    
		and~Lincheng~Shen% <-this % stops a space    Zhongkui~Li,~\IEEEmembership{Senior~Member,~IEEE,}
		\thanks{Y. Yu, X. Wang and L. Shen are with the College of Intelligence
			Science and Technology, National University of Defense Technology, Changsha,
			410073, China (e-mail: yuyangguang11@nudt.edu.cn; xkwang@nudt.edu.cn;
			lcshen@sina.com).
		}% <-this % stops a spaceAustralia. B. D. O. Anderson is also with Data61-

		\thanks{Z. Sun is with the Department of Electrical Engineering, Eindhoven University of Technology, Eindhoven, 5600 MB, Netherlands. (email: sun.zhiyong.cn@gmail.com).}}% <-this % stops a space

	% make the title area
	\maketitle

	\begin{abstract}
		In this paper, a novel design scheme is introduced to solve the  optimal control problem for  nonlinear  systems with  unsymmetrical and state-dependent input constraints.  By introducing an initial  stabilizing   control policy as the baseline of the constructed optimal control policy, we  remove  the assumption in the current study for the adaptive optimal control, that is,  the internal dynamics should  hold zero when the state of the system is in the origin.  Particularly, nonlinear control systems with partially-unknown dynamics are investigated and the procedure to acquire the corresponding optimal control policy is presented.   The stability for the closed-loop dynamics and the optimality of the obtained control policy     are both proved.   Besides,  we apply the  proposed control design framework   to solve the optimal circumnavigation problem based on the accumulative   Fisher information   for a fixed-wing unmanned aerial vehicle (UAV).   The control performance of our algorithm is compared with that of  the existing circumnavigation control policy   in a numerical simulation.
	\end{abstract}
	
	% Note that keywords are not normally used for peerreview papers.
	\section{Introduction}

	%In \cite{he2019adaptive}, the optimal controller was constructed by using the method of  linear differential inclusion. 
	
	\emph{Literature review:} 
   In the domain of automatic control, the basic requirement for the
  controller is to stabilize the system and drive the interested state to
  an equilibrium state. But when the related resource is limited or the
  system is required to compete for a specific performance index, the
  optimal behavior of the system with respect to specified long-term
  goals is desired. Therefore, optimal control for nonlinear systems has been the focus of the research
		since last century \cite{vlassenbroeck1988chebyshev} as it can help improve the system performance effectively. There have been numerous successful applications of nonlinear optimal control to different fields such as  spacecraft attitude control \cite{9559763}  and underwater vehicle  control \cite{DENG2021102676}.  In general, the optimal control problem  for  nonlinear systems involves the solving  of an underlying Hamilton-Jacobi-Bellman (HJB) equation  \cite{lewis2012optimal}, which  is usually very difficult to solve and almost impossible to get an analytical solution directly \cite{beard1998approximate}.  To solve the HJB equation, the paper \cite{abu2005nearly}  proposed an off-line policy iteration (PI) strategy, in which a sequence of cost functions were approximated.  The methods proposed in \cite{abu2005nearly}  requires  that the dynamics of the system is completely known. However, the dynamics of  the nonlinear system  is usually complex and even time-variant in some situations. As a consequence,    nonlinear systems are rather difficult to be modeled accurately. For traditional model-based control design methods, the  performance degradation  caused by  the model inaccuracy may be catastrophic. 
	
  To overcome the difficulties mentioned above, the adaptive dynamic programming (ADP) \cite{bian2014adaptive,luo2019event,9306903} method was developed. Different from the traditional model-based control design methods,  the ADP method approximates the solution of the HJB equation using  online data and further constructs the optimal control law  adaptively with the system's dynamics being partially or completely unknown  \cite{jiang2022bias,modares2014optimal,7214313}. It was proved in previous studies \cite{vamvoudakis2015asymptotically,modares2014optimal} that the ADP method can guarantee the ultimate uniform boundedness (UUB) of the system and thus the risk  of system instability caused by the  model inaccuracy can be avoided.    For systems with   partially-unknown dynamics, \cite{vrabie2009neural,Xue2021,vrabie2009adaptive} proposed the integral reinforcement learning (IRL) method to approximate the solution of the HJB equation.       When the dynamics of the system is completely unknown, an  identifier-critic-actor-based
	structure is usually used.  A  neural network (NN) \cite{modares2013adaptive} or recurrent neural network \cite{zhang2011data} was  utilized to fully identify the unknown system dynamics.  Recently, the work \cite{zhang2020deterministic} proposed a deterministic policy gradient adaptive dynamic programming   algorithm for for solving model-free optimal control problems.  Although the preconditions and the proposed  methods in the studies mentioned above are different, there exists one hidden assumption in common among these works, that is, the internal dynamics of the system should be zero when the state of the system is in the origin \cite{modares2014optimal,7214313,vamvoudakis2015asymptotically,vrabie2009neural,Xue2021,vrabie2009adaptive}.   The system that satisfies this assumption is termed as the standard form (SF) system  in this paper.  However, this assumption is not satisfied in many nonlinear systems, which are termed as    nonstandard form (NF) systems in this paper. For example, the control problems of target tracking \cite{modares2014optimal} or UAV circumnavigation  \cite{dong2019flight} involve with the NF system. It is still an unsolved problem on how to tackle the optimal control problem for the NF system. 
	
  Another important issue that is worth considering is the amplitude limitation on the control input. There often exists an unsymmetrical and state-dependent saturation zone for the input of the system's actuator in reality. Taking the attitude control of  a vehicle or aircraft  for example, the steering mechanism of a ground or aerial vehicle may partially loss effectiveness due to motor fault \cite{kong2019asymmetric}. As a consequence, the vehicle's maximum steering capacity for the left direction and the right direction may be different. Meanwhile, to avoid the risk of rollover, the maximum angular velocity of the vehicle is usually required to decrease with the increment of the linear velocity.        To confront the optimal control problem with symmetrical input constraints,  a non-quadratic  cost function was proposed in  \cite{lyshevski1998optimal} and a smooth saturated controller was further constructed.  Similar studies are reported in  the literature therein \cite{modares2014optimal,modares2013policy,Xue2021,9358462}. In these studies, the input $u$ is constrained in a symmetrical and fixed set, i.e.,  $u$ is constrained by $|u|<\lambda$ with $\lambda$ being a positive constant.   For the optimal control problem with unsymmetrical input constraint,  the paper \cite{zhou2018neuro}  proposed an ADP-based neuro-optimal controller for discrete-time nonlinear systems with asymmetric input saturation. More recently, the work \cite{kong2019asymmetric} proposed an adaptive optimal control law by introducing a switching function.  We notice that the switching function in \cite{kong2019asymmetric} should be carefully selected to guarantee the stability of the closed-loop system.   The event-triggered adaptive optimal control problem was studied    in \cite{8472160} and \cite{9032344} for a class of asymmetrically input-constrained nonlinear systems. The work \cite{yang2020optimal} constructed an optimal neurocontroller under the framework of RL and the work  \cite{9013018} presented an event-driven $H_\infty$ controller for continuous nonlinear systems with asymmetric
  input saturation.  However,  the work  \cite{8472160,9032344,yang2020optimal,9013018}  can only guarantee the UUB of the closed-loop system theoretically.  In summary,  the proposed methods in the existing studies still exist some shortcomings.  Meanwhile, none of these works mentioned above considered the adaptive optimal control problem with state-dependent input constraint.  
	
	To demonstrate the application value of the method proposed in this paper, we apply  the proposed algorithm to solve the optimal UAV circumnavigation control problem, which is another main contribution of this paper. Although there have been numerous applications of UAV, the surveillance and tracking of moving ground targets is still one of the most important applications  of UAV  \cite{liu2019mission}.  To monitor a ground tar circumnavigate  around the ground target with a preset radius \cite{sun2018collaborative}.  For robots with single-integrator dynamics, different control algorithms have been proposed  to achieve the circumnavigation with distance measurements \cite{shames2011circumnavigation} or bearing measurements \cite{deghat2014localization}. For a non-holonomic agent,  \cite{deghat2012target} proposed a circumnavigation control method while  the position of the target is assumed to be unknown.   Assuming the relative position of the target is accessible, the paper  \cite{dong2019flight} designed a guidance law by exploiting the Vector Fields (VF) method.  However, none of these works have considered  the optimality  issue of the circumnavigation control. 
	
%	Following their works,  \cite{cao2015uav}  designed a sliding-mode estimator of the range rate and further developed a control strategy  for a UAV with only  range measurements.
	
	%The primary objective of this paper is to develop a set of controller design approaches for optimal control in  nonlinear control affine systems with  unsymmetrical input constraints.
	\emph{Statement of contributions:}   Aimed at solving the optimal control problem for nonlinear systems with  state-dependent and unsymmetrical input constraints, this paper   investigates the partially-unknown system whose dynamics is in NF form and an online PI algorithm is presented in this paper. The main contributions of this paper are summarized as follows:
 
	\begin{itemize}
	 	\item[1)]  An online PI algorithm is proposed to address the optimal control of nonlinear  systems with state-dependent and unsymmetrical input constraints. The stability and convergence of the proposed algorithm are also proved.  Compared with the existing studies, our method has the following novelties:
			\begin{itemize}
				\item  Compared with the existing studies such as 	\cite{modares2014optimal,lyshevski1998optimal,modares2013policy,9358462,kong2019asymmetric,yang2020optimal}, this paper  addresses the optimal control problem with the constraint set of the input $u$ being state-dependent rather than being constant. 
				\item Compared with \cite{8472160,yang2020optimal,9013018}, the closed-loop system with the proposed algorithm  is theoretically guaranteed to be asymptotically stable rather than being UUB. Moreover, the  method proposed in this paper does not exist the difficulty of designing    switching functions in comparison with \cite{kong2019asymmetric}.
%				\item  Different from the PI-based ADP algorithm proposed in the previous studies such as \cite{liu2013adaptive,vrabie2009neural}, the initial weight of the neural network can be directly set as zero. In \cite{liu2013adaptive,vrabie2009neural}, the initial NN weight  should be carefully selected so that the corresponding control law based on the neural network can stabilize the system, which is difficult for most nonlinear systems.
		\end{itemize}  
 		
			\item[2)] The optimal control problem for a NF system is addressed. In the current study of  the   adaptive optimal control, the internal dynamics is usually required to hold zero when the state of the system is in the origin. To expand the application range of  the   ADP theory to    systems that do not meet this condition, this paper designs a special control law which consists of the initial stabilizing control law and the neural-network-based control law.	
		\item[3)] By  exploiting the method proposed in this paper, an  optimal circumnavigation control law with input saturation is proposed. To the best of our knowledge, it is the first time that the  UAV optimal circumnavigation problem w.r.t. an  infinite-horizon performance index is addressed.	 		
	\end{itemize}	

	The rest of the paper is organized as follows:  Section~\ref{sec:4} develops our  adaptive optimal control algorithm for the nonlinear   systems with state-dependent and unsymmetrical input constraints.    In Section~\ref{sec:5},  we apply our methods to solve  the optimal UAV circumnavigation control problem and a comparison with the method  proposed in  \cite{dong2019flight} is illustrated.  Finally, the concluding remarks are drawn in Section~\ref{sec:6}.

	\textbf{Notation:} The vector $\boldsymbol{1}_n\in\mathbb{R}^n$ denotes a   vector with its elements all being 1 and $\boldsymbol{0}_{n\times m} \in \mathbb{R}^{n\times m}$ is a zero matrix.    Here we define an operator $\boldsymbol{vec}(\cdot):\mathbb{R}^{n\times n} \rightarrow \mathbb{R}^n$. For a vector $\boldsymbol{a}\in\mathbb{R}^n$ and a diagonal matrix $M\in\mathbb{R}^{n\times n}$,  if $\boldsymbol{a}= \boldsymbol{vec}(M)$, one has $a_i =M_{ii}, i=1,\cdots,n$, where $a_i$  is the $i$-th element of  the vector $\boldsymbol{a}$ and $M_{ii}$ is the $i$-th element on the diagonal of the matrix $M$.
	
	%\section{Mathematical preliminaries} \label{sec:2}

	\section{The   optimal control problem for NF systems with  unsymmetrical input constraints }
	%In this section, we give the formulation of the standard optimal control problem for systems with input constraints.
	
	%In this section, we firstly present the process of designing an information optimal circumnavigation control law. A neural network is then used to approximate the optimal control law and the method of tunning the weights of the network is also introduced.  

	%\subsection{Neural-network-based policy iteration algorithm for solving the HJB equation}
	\label{sec:4}
	\subsection{Problem formulation} \label{sec:method}
	Consider the following system whose dynamics
	is   	
	\begin{numcases}{}
	\dot{x}_1(t)=f_1(x_1,x_2)+g_1(x_1,x_2)u(t), \label{eq:sys2} \\
	\dot{x}_2(t)=f_2(x_2), \label{eq:sysx2}
	\end{numcases}
	where $x_1 \in \mathbb{R}^{n_1}$ is the system state  to be stabilized; $x_2 \in \mathbb{R}^{n_2}$ is the state which is not intended to be controlled and assumed to be bounded; the continuous  functions $f_1(x_1,x_2) \in \mathbb{R}^{n_1}$ and $f_2(x_2) \in \mathbb{R}^{n_2}$ are the unknown internal dynamics of the system; $g_1(x_1,x_2) \in \mathbb{R}^{n_1 \times m}$  is  the input dynamics of the system;  $u(t) \in \mathbb{R}^{m}$ is the control input. Note that the function $f_1(x_1,x_2)$ is not necessary to be zero when $x_1=0$.
	
	Denote    the stack vector $x\in\mathbb{R}^n$ as $x=[x_1^\top,x_2^\top]^\top$, where $n=n_1+n_2$. The control input $u(t)$ is constrained by the condition 
	\begin{equation}
	d_i(x(t))\le u_i(t) \le  h_i(x(t)), \ i=1,\cdots,m, \label{eq:ulimit}
	\end{equation}
	where   $u_i(t)$ is the $i$-th element of $u(t)$;  $d_i(x(t))$ and $ h_i(x(t))$ are the known functions that determine the lower bound and the upper bound for the $i$-th element of $u(t)$.   
	%Denote the stack vectors of the input's lower bound and upper bound as  $\overline{D}(t)=[d_1(t),d_2(t),\cdots, d_m(t)]^\top$  and  $\overline{H}(t)=[h_1(t),h_2(t),\cdots, h_m(t)]^\top$, respectively.  Then \eqref{eq:ulimit} can be rewritten in the following concise form:
	%\begin{equation}
	%\overline{D}(t) \le u(t) \le \overline{H}(t). \notag
	%\end{equation}
	
	\begin{remark}
The system described by \eqref{eq:sys2} and \eqref{eq:sysx2} can be regarded as the general form of many widely studied systems. For example, the state $x_2$  can be regarded as a bounded time-varying uncertainty \cite{sun2016gpio} and  system  \eqref{eq:sys2} is  to be stabilized while disturbed by $x_2$. Also, the system described by \eqref{eq:sys2} and \eqref{eq:sysx2}  is commonly used in the target tracking system with  $x_1$ being the tracking error and $x_2$ being the state of the target. There have been many works reported on such systems in the  previous literatures,  such as \cite{modares2014optimal,KUMAR2021109558}. 
	\end{remark}
	\begin{remark}
		In this paper, the constraint set of $u$ is unsymmetrical and state-dependent as \eqref{eq:ulimit} reveals. Meanwhile, the internal dynamics $f_1(x_1,x_2)$ of   system  \eqref{eq:sys2} does not satisfy the condition that $f(x_1,x_2)=0$ when $x_1=0$, which is required in the previous studies of ADP  \cite{modares2014optimal,7214313,vamvoudakis2015asymptotically,vrabie2009neural,Xue2021,vrabie2009adaptive,modares2013policy}. As a consequence, the  methods proposed in  the  previous studies, such as \cite{Xue2021,modares2013policy},  are inappropriate. 
	\end{remark}

	The aim of this paper is to design an optimal  policy $u^*(t)=\mu^*(x)$  constrained by the unsymmetrical set \eqref{eq:ulimit} such that  the  $x_1$-system is stabilized as well as    a  performance index $	\mathcal{J}(x(0), u)$ defined in the following form is minimized:
\begin{equation}
	\mathcal{J}(x(0), u)=\int_{0}^{\infty} \left[Q(x_1)+U_n(u,x)\right] \mathrm{d}\tau,  \  x(0)=x^o,   \label{eq:J2}
\end{equation}
where $Q(x_1)$ is a positive semi-definite  function related with the state $x_1$,  $U_n(u,x)$ is a positive semi-definite function which is to be designed, and $x^o$ is the initial state.  

	Before presenting the solution to the optimal control problem  described above,  we  firstly introduce a definition of the admissible control.

 \begin{Def}[Admissible Control  \cite{abu2005nearly}]
		For a given system described by \eqref{eq:sys2} and \eqref{eq:sysx2}, a control  policy $u(t)=\mu(x)$ is defined to be  admissible with respect to the  performance index \eqref{eq:J2} on a compact set $\Omega \subseteq \mathbb{R}^n$, written as $\mu(x) \in \mathcal{A}(\Omega)$, if $\mu(x)$ is continuous, $u(t)=\mu(x)$ stabilizes   system \eqref{eq:sys2} and $\mathcal{J}(x^o,u)$ is finite for every $x^o\in \Omega$. \label{def:admissible}
\end{Def} 
	
	Similar to the previous works, such as  \cite{modares2014optimal,vrabie2009neural}, the following assumption is set in this paper.

	\begin{assump}
		There exists a known admissible  control policy on a set $\Omega \subseteq \mathbb{R}^{n}$  which stabilizes system \eqref{eq:sys2} and satisfies    constraint \eqref{eq:ulimit}. \label{assump2}
	\end{assump}

		\begin{remark}
		  In many scenarios of nonlinear system controls with input saturation, such as trajectory-tracking control \cite{wei2010stabilization} and formation control \cite{yang2021event}, some Lyapunov-based methods have been proposed to design a stabilizing but non-optimal control policy with input saturation, and thus an initial admissible control policy can be obtained.  In some industrial applications, the initial admissible controller also can be constructed by empirical methods, such as tunning the control parameters of a PID controller.   
		\end{remark}

%		Although the  optimal control problem for systems with input constraints has been addressed, the following two preconditions are required for the validity of the method  reviewed above:
	
%	\noindent(\romannumeral1) The internal dynamics of the system $f(x)$  has to satisfy the condition $f(0)=0$;
%	
%	\noindent (\romannumeral2) The input constraint has to be a symmetrical and fixed set, i.e., $|u_i(t)|\le \lambda, i=1,\cdots,m $.
%	
%	In the next section, we will illustrate how to design an optimal control policy for nonlinear systems which do not satisfy the preconditions listed above. 
%	

	\subsection{The design method for NF systems with unsymmetrical and state-dependent input constraints } \label{sec:design}  
If Assumption~\ref{assump2} holds,  the initial admissible control policy is denoted as $u_s(t)=\mu_s(x(t))$.  	Although the initial control policy $u_s(t)$ is stabilizing, the control performance of $u_s(t)$ may not be satisfactory.  Thus, based on the initial control policy $u_s(t)$, we design  the control policy $u(t)$ as
	\begin{equation}
	u(t)=u_s(t)+\hat{u}(t), \label{eq_uus}
	\end{equation}  
	where $\hat{u}(t)$ is a to-be-designed virtual input which enables the actual control input $u(t)$  to achieve the optimal control performance.  As $u(t)$ is constrained by $\eqref{eq:ulimit}$, the virtual input  $\hat{u}(t)$  should satisfy
	\begin{equation}
	d_i(x(t))-u_s^i(t)\le\hat u_i(t)\le  h_i(x(t))-u_s^i(t), \ i=1,\cdots,m, \label{hatuilimit}
	\end{equation}
	where $\hat{u}_i(t)$ and $u_s^i(t)$ are the $i$-th element of $\hat{u}(t)$ and $u_s(t)$, respectively.  Further by employing \eqref{eq_uus}, system \eqref{eq:sys2} is rewritten as
	\begin{equation}
	\dot{x}_1(t)=F_s(x_1,x_2)+g_1(x_1,x_2)\hat{u}(t),\label{eq_sysvery}
	\end{equation}
	where the function $F_s(x_1,x_2)$ is  defined by
	\begin{equation}
	F_s(x_1,x_2)=f_1(x_1,x_2)+g_1(x_1,x_2)\mu_s(x). \notag
	\end{equation}

	Define two sets of functions $\overline{\lambda}_i(u,x)$ and $\hat{\lambda}_i(\hat{u},x)$, $i=1,2,\cdots,m$, as
	\begin{align}
	\overline{\lambda}_i(u,x)&=\left\{\begin{array}{lll}
	h_i(x)-\mu_s^i(x),    \   &\text{if}  &u_i-\mu_s^i(x)\ge 0, \\
	- d_i(x)+\mu_s^i(x),   \   &\text{if}  &u_i-\mu_s^i(x) < 0,
	\end{array}\right. \notag \\
	\hat{\lambda}_i(\hat{u},x)&=\left\{\begin{array}{lll}
	h_i(x)-\mu_s^i(x),    \   &\text{if}  &\hat{u}_i\ge 0, \\
	- d_i(x)+\mu_s^i(x),   \   &\text{if}  &\hat{u}_i < 0.
	\end{array}\right. \label{eq:overlamda}
	\end{align}

  Note that if Assumption~1 is satisfied, it holds that
	\begin{equation}
	d_i(x)< \mu_s^i(x) < h_i(x), i=1,\cdots,m,  \notag
	\end{equation}
	which yields 	
	\begin{equation}
	d_i(x)-\mu_s^i(x)< 0 < h_i(x)-\mu_s^i(x), i=1,\cdots,m, 	\label{assum:ulimit}
	\end{equation}
	where $\mu_s^i(x)$ is the $i$-th element of the function $\mu_s(x)$. The inequality \eqref{assum:ulimit}  guarantees that $\overline{\lambda}_i(u,x)=	\hat{\lambda}_i(\hat{u},x)\ge 0$ if $ \hat{u}_i=u_i-\mu_s^i(x)$. 
	
Inspired by the previous studies such as \cite{modares2014optimal},  we design the control cost function $U_n(u,x)$ for   system \eqref{eq:sys2} as
	\begin{align}
	U_n(u,x)=&2 \sum_{i=1}^{m}\int_{0}^{u_i-\mu_s^i(x)}\overline{\lambda}_i r_i \left(\tanh^{-1}(s/\overline{\lambda}_i)\right) \mathrm{d} s, \label{eqUndefi}
	\end{align}
	where  $r_i$ is a positive constant and $\overline{\lambda}_i \triangleq \overline{\lambda}_i(u,x)$.
	\begin{remark}
		 Note that compared with the design of the previous literatures, such as \cite{modares2014optimal,modares2013policy,Xue2021}, the scaling factor $\overline{\lambda}_i $ is a function related with the system state rather than a constant in order to tackle the unsymmetrical input constraint and the NF form of the system. Meanwhile, the upper bound of the integral in \eqref{eqUndefi} takes the initial admissible control policy $\mu_s(x)$ into consideration, which is also different from the design of the previous studies.
	\end{remark}

The issue that remains is to  design a virtual input $\hat{u}(t)$ such that the control policy $u(t)$ given by \eqref{eq_uus} is optimal w.r.t.  the performance index \eqref{eq:J2}. To achieve this target, we define a performance index $\mathcal{J}_2(x(0),\hat{u})$ for system \eqref{eq_sysvery} as
	\begin{equation}
	\mathcal{J}_2(x(0),\hat{u})=\int_{0}^{\infty} [ Q(x_1)+\hat{U}_n(\hat{u},x)  ]\mathrm{d} \label{eq:J3}\tau, \   x(0)=x^o, 
	\end{equation}
	where the function $\hat{U}_n(\hat{u},x)$ is defined by
	\begin{align}
	\hat{U}_n(\hat{u},x)=&2 \sum_{i=1}^{m}\int_{0}^{\hat{u}_i}r_i\hat{\lambda}_i  \left(\tanh^{-1}(s/\hat{\lambda}_i )\right) \mathrm{d} s \notag \\
	=&2 \sum_{i=1}^{m} \hat{\lambda}_i \tanh ^{-1}\left( \hat{u}_i   / \hat{\lambda}_i\right)  r_i \hat{u}_i  \notag \\
	&+  \sum_{i=1}^{m}\hat{\lambda}_i^{2} r_i \ln \left(\boldsymbol{1}_n-\left(\hat{u}_i / \hat{\lambda}_i\right)^{2}\right),
	\label{eq:Wudesigned}
	\end{align}
	and $\hat{\lambda}_i\triangleq \hat{\lambda}_i(\hat{u},x)$.  Based on the design of the functions $\overline{\lambda}_i(u,x)$, $\hat{\lambda}_i(\hat{u},x)$, $U_n(u,x)$ and  $\hat{U}_n(\hat{u},x)$ in \eqref{eq:overlamda}, \eqref{eqUndefi} and \eqref{eq:Wudesigned}, we will show how to design an optimal control law for NF systems with unsymmetrical and state-dependent input constraints in the following part.  
	
	 	The function $F_s(x_1,x_2)$ and the virtual input $\hat{u}(t)$ can be regarded as the internal dynamics and control input of system \eqref{eq_sysvery}, respectively. As the initial control policy $u_s(t)$ is an admissible control of system \eqref{eq:sys2}, it holds that $F_s(0,x_2)=0$. Otherwise, when $u(t)=u_s(t)$ and $x_1=0$, it holds that $\dot{x}_1=F_s(0,x_2)\neq 0$, which contradicts with the fact that system \eqref{eq:sys2} can be stabilized by  the control policy $u_s(t)$. Moreover, since $F_s(0,x_2)=0$, the virtual input $\hat{u}$ should hold zero when $x_1=0$. 	 As a consequence, by using  \eqref{eq_uus},  system \eqref{eq:sys2}  is transformed into system    \eqref{eq_sysvery}, which meets the requirement for the admissible control defined in Definition~\ref{def:admissible}. 
	
		\begin{remark}
		In the previous works, such as \cite{modares2014optimal,vrabie2009neural},  the initial admissible control policy  $u_s(t)$ is only used in the initial policy iteration.  However, in this paper, it can be observed from \eqref{eq_uus} that the initial admissible control policy  $u_s(t)$ also acts as a baseline for constructing the optimal control law. Then the  to-be-designed virtual input $\hat{u}(t)$ is   added on the initial control policy $u_s(t)$ to obtain an optimized control performance.   By this method, the internal dynamics $f_1(x_1,x_2)$ of system \eqref{eq:sys2} is not necessary to be zero when $x_1=0$.  
	\end{remark}

	The following lemma shows that the optimal control problem  for  system  \eqref{eq:sys2} can be transformed  into the optimal control problem for  system  \eqref{eq_sysvery}.
	\begin{lemma}
		Given a    unique  optimal control policy  $\hat{u}^*(t)$  for system \eqref{eq_sysvery} w.r.t. the performance index \eqref{eq:J3},   the control policy $u^*(t)$ given by
		\begin{equation}
		u^*(t)=\hat{u}^*(t)+u_s(t) \label{eq:invu}
		\end{equation}
		is  the unique optimal control policy  for  system \eqref{eq:sys2} w.r.t. the performance index \eqref{eq:J2}. \label{lemma2}
	\end{lemma}
	\begin{IEEEproof}
		Given two control policies $u(t)$ and $\hat{u}(t)$ satisfying \eqref{eq_uus} with the same initial states,  the trajectories of system \eqref{eq:sys2} and  system \eqref{eq_sysvery} are identical. Meanwhile, from 	\eqref{eq:overlamda}, it can be observed that $\overline{\lambda}_i(u,x)=\hat{\lambda}_i(\hat{u},x)$. Further, from \eqref{eqUndefi} and \eqref{eq:Wudesigned}, it holds that $U_n(u,x)=\hat{U}_n(\hat{u},x)$ if $u=\mu_s(x)+\hat{u}$, which further implies $\mathcal{J}(x(0), u)=\mathcal{J}_2(x(0), \hat{u})$. 
		Suppose the control policy $u^*(t)$ given by \eqref{eq:invu} is not the unique optimal control policy, there always exists a control policy $u'(t)$ such that
		\begin{equation}
		\mathcal{J}(x(0), u')\le \mathcal{J}(x(0), u^*). \label{eqJlehatJ}
		\end{equation}
		Define a control policy  $\hat{u}'(t)=u'(t)-u_s(t)$. 
		Through the discussion above, one has
		\begin{equation}
		\mathcal{J}(x(0), u')=\mathcal{J}_2(x(0), \hat{u}'). \label{eqJlehatJ2}
		\end{equation}
		By using \eqref{eqJlehatJ} and \eqref{eqJlehatJ2}, it yields
		\begin{equation}
		\mathcal{J}_2(x(0), \hat{u}')=\mathcal{J}(x(0), u')\le\mathcal{J}(x(0), u^*)=\mathcal{J}_2(x(0), \hat{u}^*), \notag
		\end{equation}
		which contradicts with the fact that  $\hat{u}^*(t)$ is the optimal control policy. Thus, $u^*(t)$ given by \eqref{eq:invu}  is  the unique optimal control policy  for  system \eqref{eq:sys2} w.r.t. the performance index \eqref{eq:J2}.
	\end{IEEEproof}

	Lemma~\ref{lemma2} shows that the optimal control policy $u^*(t)$  for  system \eqref{eq:sys2} w.r.t. the performance index \eqref{eq:J2} can be obtained if the optimal control policy $\hat{u}^*(t)$ is available. Thus, we introduce the  procedure of designing  $\hat{u}^*(t)$ in the following part.

	Using  \eqref{eq:sysx2} and \eqref{eq_sysvery},  the dynamics of the  state $x$ can be rewritten as
	\begin{equation}
	\dot{x}=F(x)+G(x)\hat{u}, \label{eq:totalsys}
	\end{equation}
	where
	\begin{align}
	F(x)&=[F_s(x)^\top,f_2(x_2)^\top]^\top, \notag \\
	G(x)&=[g_1(x_1,x_2)^\top,\boldsymbol{0}_{n_2\times m}^\top]^\top. \notag
	\end{align}

	Assume there exists a continuously differentiable   value function $V^*(x)$ defined by
	\begin{equation}
	V^{*}(x(t))= \min _{\hat{u} \in \mathcal{A}(\Omega)} \int_{t}^{\infty}\left[ Q(x_1)+\hat{U}_n(\hat{u},x) \right]\mathrm{d} \tau. \label{eqVstar}
	\end{equation}
	A Hamiltonian function $\hat{H}(x,V^*,\hat{u})$ is defined  as
	\begin{align}
	\hat{H}(x,V^*,\hat{u})=&(V_x^*)^\top(F+G\hat{u})+Q(x_1) +\hat{U}_n(\hat{u},x), \label{eq:Halmiton}
	\end{align}
	where $F\triangleq F(x)$,  $G\triangleq G(x)$, $V^* \triangleq V^*(x)$,  and $ V_x^* = \partial V^*(x)/\partial x \in \mathbb{R}^{n}$.  Then,  similar to the previous works \cite{modares2014optimal,modares2013policy,Xue2021},  by using the stationarity condition (see \cite{lewis2012optimal}) on the Hamiltonian function  $\hat{H}(x,V^*,\hat{u})$, i.e., $\partial \hat{H}/\partial \hat{u}=0$, we obtain the optimal control policy $\hat{u}^*(t)$ as
	\begin{equation}
	\begin{aligned} \hat{u}^{*}(t) &=\hat{\mu}^*(x(t))=\underset{\hat{u} \in \mathcal{A}(\Omega)}{\arg \min } \hat{H}\left(x,V^*,\hat{u}\right) \\
	&= -\hat{\lambda} \tanh \left( (1/2) (\hat{\lambda}R)^{-1}G^{\top}  V_x^*\right), \label{eq:ustar1}
	\end{aligned}
	\end{equation}
	where $\hat{\lambda},R \in \mathbb{R}^{m\times m}$ are diagonal matrices whose $i$-th elements on the diagonal are $\hat{\lambda}_i$ and $r_i$, respectively.   It can be observed from \eqref{eq:ustar1} that $|\hat{u}_i|\le \hat{\lambda}_i(\hat{u}_i,x)$, based on which the inequality \eqref{hatuilimit} can be derived.  Further,  the control policy $u(t)$ defined by \eqref{eq_uus} can satisfy the constraint \eqref{eq:ulimit}.  
	
	Next, substituting \eqref{eq:ustar1} into \eqref{eq:Wudesigned} results in
	\begin{align}\hat{U}_n\left(\hat{u}^{*},x\right)=&  ( V_x^{*})^{ \top} G \hat{\lambda}\tanh (\hat{D}^{*} ) \notag \\
	&+ (\boldsymbol{vec}(\hat{\lambda}R\hat{\lambda}))^\top \ln \left(\boldsymbol{1}_n-\tanh ^{2} (\hat{D}^{*})\right), \label{eq:expUn}
	\end{align}
		where $\hat{D}^*=(1 / 2) (\hat{\lambda}R)^{-1}  G^{\top} V_x^*$. By putting  \eqref{eq:expUn}  and \eqref{eq:ustar1} into  \eqref{eq:Halmiton},   it yields the following HJB equation:
	\begin{align}
	(\boldsymbol{vec}(\hat{\lambda}R\hat{\lambda}))^\top&\ln \left(\boldsymbol{1}_n-\tanh^2(\hat{D}^*)\right) \notag \\
	&+Q(x_1)+   (V_x^*)^\top F(x) =0\label{eq:Hal_eq2}.
	\end{align}
	
	If the solution $V_x^*$ of the HJB equation \eqref{eq:Hal_eq2} is found, the  optimal virtual  input $\hat{u}^*$ for  the system \eqref{eq_sysvery} can be obtained by \eqref{eq:ustar1}, and the optimal control policy  $u^*(t)$ for system \eqref{eq:sys2} is further obtained according to \eqref{eq:invu}.
	
	In the following theorem, it is proved that the control law $u^*(t)$ defined by \eqref{eq:invu}  and \eqref{eq:ustar1} is the unique optimal control policy w.r.t. the performance index \eqref{eq:J2} and stabilizes the  $x_1$-system.
	
	\begin{theorem}
		Consider the optimal control problem for system \eqref{eq:sys2} w.r.t. the performance index \eqref{eq:J2}. Suppose that $V^*(x)$  is a   positive-definite solution to the  HJB equation \eqref{eq:Hal_eq2}. Then the control policy $u^*(t)$ defined by \eqref{eq:invu}  and \eqref{eq:ustar1} is  the unique optimal control policy  such that  system \eqref{eq:sys2} is stabilized asymptotically  and  the performance index \eqref{eq:J2} is minimized.  \label{thorem1}
	\end{theorem}
	
	\begin{IEEEproof}
		Firstly, we will prove that $u^*(t)$ is the   unique optimal policy that minimizes the performance index \eqref{eq:J2}. It has been proved in  Lemma~\ref{lemma2} that the optimality of the control policy $u^*(t)$ for system \eqref{eq:sys2} w.r.t. the performance index \eqref{eq:J2} is equivalent to   the optimality of the control policy $\hat{u}^*(t)$  w.r.t. the performance index \eqref{eq:J3}.  Thus, one only needs to prove  that  $\hat{u}^*(t)$  is  the unique optimal control policy for system \eqref{eq_sysvery} w.r.t. the performance index \eqref{eq:J3}.

		Note that for the optimal value function $V^*(x(t))$ defined by \eqref{eqVstar}, one has
		\begin{align}
		\int_{0}^{\infty} \left( \dot{V}^*(x(\tau))\right) \mathrm{d} \tau&=-V^*(x(0)). \notag
		\end{align}
		Then given an admissible  control law $\hat{u}(t)$, the performance index \eqref{eq:J3} can be rewritten as	 
			\begin{align}
			\mathcal{J}_2&(x(0), \hat{u})= \int_{0}^{\infty} \left[Q(x_1)+\hat{U}_n(\hat{u},x)\right] \mathrm{d} \tau +V^*(x(0)) \notag \\
			& +  \int_{0}^{\infty} \left( \dot{V}^*(x)\right) \mathrm{d} \tau \notag \\
			=& \int_{0}^{\infty} \left[Q(x_1)+\hat{U}_n(\hat{u},x)\right] \mathrm{d} \tau +V^*(x(0)) \notag \\
			&+\int_{0}^{\infty} \big[ (V_x^*)^{\top}  (F+G \hat{u}^*)+(V_x^*)^{\top}  G(\hat{u}-\hat{u}^*)  \big]\mathrm{d}\tau.      \label{eq:VxHJB2} 
			\end{align}
			By adding and subtracting the term  $\hat{U}_n(\hat{u}^*,x)$ in the integral part, \eqref{eq:VxHJB2} becomes
			\begin{align}
			&\mathcal{J}_2(x(0), \hat{u})=V^*(x(0)) \notag \\
			&+\int_{0}^{\infty} \big[\hat{U}_n(\hat{u},x)-\hat{U}_n(\hat{u}^*,x)+(V_x^*)^{\top}  G(\hat{u}-\hat{u}^*)  \big]\mathrm{d}\tau \notag   \\ 
			&+\int_{0}^{\infty} \left[ (V_x^*)^{\top}  (F+G \hat{u}^*)+Q(x_1)+\hat{U}_n(\hat{u}^*,x)\right] \mathrm{d} \tau. 
			\label{eq:VXHJB} 
			\end{align}

		It can be observed  from \eqref{eq:Wudesigned} and \eqref{eq:ustar1} that 
		\begin{align}
		&	\hat{U}_n(\hat{u},x)-\hat{U}_n(\hat{u}^*,x) \notag \\
		&	= 2 \sum_{i=1}^{m}\int_{\hat{u}_i^*}^{\hat{u}_i}r_i\hat{\lambda}_i  \left(\tanh^{-1}(s/\hat{\lambda}_i)\right) \mathrm{d} s. \notag \\
		&	(V_x^{*})^\top G= -2\sum_{i=1}^{m}r_i\hat{\lambda }_i \tanh^{-1}(\hat{u}_i^{*}/\hat{\lambda}_i).\notag
		\end{align}

		Thus we have
		\begin{align}
		M_u \triangleq  &\hat{U}_n(\hat{u},x)-\hat{U}_n(\hat{u}^*,x)+(V_x^*)^{\top}  G(\hat{u}-\hat{u}^*)  \notag \\
		=&2\sum_{i=1}^{m}\int_{\hat{u}_i^*}^{\hat{u}_i}r_i\hat{\lambda}_i \left(\tanh^{-1}(s/\hat{\lambda}_i)\right) \mathrm{d} s  \notag \\
		&-2\sum_{i=1}^{m}r_i\hat{\lambda }_i \tanh^{-1}(\hat{u}_i^{*}/\hat{\lambda}_i)\left(\hat{u}_i-\hat{u}_i^*\right). 
		\label{eq:M} 
		\end{align}
		Using Leibniz's rule to differentiate $V^*(x)$  along the trajectory of  system $\dot{x}=F(x)+G(x)\hat{u}^*$, it yields
		\begin{align}
		(V_x^*)^\top (F+G\hat{u}^*)=-Q(x_1) -\hat{U}_n(\hat{u}^*,x),\notag
		\end{align}	
		which implies 
		\begin{equation}
		\hat{H}(x, V^*,\hat{u}^*)=0. \label{Hstarzero}
		\end{equation}
		By employing \eqref{eq:M} and \eqref{Hstarzero},   \eqref{eq:VXHJB} becomes 
		\begin{align}
		\mathcal{J}_2(x(0),  \hat{u})=&\int_{0}^{\infty} \hat{H}(x, V^*,\hat{u}^*) \mathrm{d} \tau + \int_{0}^{\infty}M_u(\tau) \mathrm{d}\tau \notag \\
		&+V^*(x(0)) \notag \\	
		=&\int_{0}^{\infty}M_u(\tau) \mathrm{d}\tau +V^*(x(0)).  \label{eq34}
		\end{align}	
		
 	To prove that $\hat{u}^*(t)$  is the unique optimal control policy, one only needs to show that $M_u>0$ always holds for all
			$\hat{u}\neq \hat{u}^*$ and $M_u$ equals   zero if and only if $\hat{u}=\hat{u}^*$. Here we define a function $f_\alpha(a,b)$ as 
			\begin{equation}
			f_\alpha(a,b)=\int_{a}^{b}\beta(s)\mathrm{d}s-\beta(a)(b-a), \label{eq:falpha}
			\end{equation}
			where $\beta(s)$ is a monotonically increasing function. It can be verified that $f_\alpha(a,b)> 0$ always holds for $\forall a\neq b$ and $f_\alpha(a,b)= 0$ when $a=b$. Let $\beta(s)=\tanh^{-1}(s/\hat{\lambda}_i)$ and 	for all $\hat{u}\neq \hat{u}^*$, we have 
			\begin{equation}
			M_u=2r_i\hat{\lambda }_i\sum_{i=1}^{m}f_\alpha(\hat{u}_i^{*},\hat{u}_i )> 0.\notag
			\end{equation}

	 Note that $V^*(x)$ is a positive semi-definite function  and  $V^*(x)=0$ if and only if $\lVert x_1\lVert=0$. As a consequence, the function $V^*(x)$ can be utilized as a Lyapunov function for $x_1$.  From $H(x, V^*, \hat{u}^*)=0$, it yields
		\begin{align}
		\frac{d V^*(x)}{dt}&=(V_x^*)^{\top}  (F+G \hat{u}^*(x)) \notag \\
		&=-Q(x_1)-\hat{U}_n(\hat{u},x)\le 0. \label{eq:dvX}
		\end{align}
		The equality in \eqref{eq:dvX} holds if and only if $\lVert x_1 \rVert=0$. As a consequence, the $x_1$-system is asymptotically stable.
	\end{IEEEproof}
	
	%\begin{remark}
	%	In this paper, the dynamics of system is assumed to be partially unknown. However, it is not difficult to observe that the method described in this section is a design scheme. Combined with the methods proposed in  the  previous literatures like \cite{abu2005nearly,lv2016online}, our method can still work when the dynamics of system is completely known or unknown. 
	%\end{remark}
	
	\subsection{Online policy iteration algorithm for solving the HJB-equation} \label{sec:HJB}
	 To derive $\hat{u}^*(t)$  based on \eqref{eq:ustar1}, the solution $V_x^*$ of the HJB equation \eqref{eq:Hal_eq2} has to be solved.    However, since \eqref{eq:Hal_eq2} is usually highly nonlinear, it is quite difficult to get its   analytical solution. In the following part,  to obtain  an equivalent formulation of HJB which does not need the knowledge of the internal dynamics $f_1(x_1,x_2)$ and $f_2(x_2)$,  the IRL idea introduced in \cite{vrabie2009adaptive} is employed.  
	
	Let   $T > 0$ denote the integral reinforcement interval and it holds that
	\begin{align}
	\begin{aligned} V(x(t))= \int_{t}^{t+T}  &\big[ Q( x_1) +\hat{U}_n(\hat{u},x) \big] \mathrm{d} \tau \\
	&+ V(x(t+T)).
	\end{aligned} \label{eq:IRLBell}
	\end{align}
	Based on \eqref{eq:IRLBell}, the following IRL-based PI algorithm is utilized to get the solution of the HJB equation \eqref{eq:Hal_eq2} with  the internal dynamics $f_1(x_1,x_2)$ and $f_2(x_2)$ unknown.
	
	1. (Policy evaluation) given an admissible control policy $\hat{\mu}^{(k)}(x)$, update $V^{(k)}(x)$ by
	the Bellman equation
	\begin{align}
	\begin{aligned}
	V^{(k)}(x(t))=\int_{t}^{t+T} &\big[ Q( x_1) +\hat{U}_n(\hat{u},x) \big] \mathrm{d} \tau  \\
	&+ V^{(k)}(x(t+T)).   
	\end{aligned}\label{eq:IRLBell2}
	\end{align}
	2. (Policy improvement) update the control policy according to

	\begin{equation}
	\hat{\mu}^{(k+1)}(x)= -\hat{\lambda} \tanh \left( (1/2) (\hat{\lambda}R)^{-1}G^{\top}  (x) V_x^{(k)}\right), \label{eq:pL}
	\end{equation}
	where  $V_x^{(k)}\triangleq \partial V^{(k)}(x) / \partial x$; the notations  $V^{(k)}(x)$ and $\hat{\mu}^{(k)}(x)$ represent the value function and the virtual control policy in the $k$-th iteration, respectively.

	The  following theorem shows that the IRL method introduced above can be employed to improve the control law.

	\begin{theorem}
		Let $\hat{u}^{(k)}=\hat{\mu}^{(k)}(x)\in \mathcal{A}(\Omega)$ and  $V^{(k)}(x)$ satisfy  $H(x,V^{(k)},\hat{u}^{(k)})=0$ 	
		with the boundary condition $V^{(k)}(0)=0$. Then, the control policy  $\hat{\mu}^{(k+1)}(x)$ defined by \eqref{eq:pL} is an admissible control for  system \eqref{eq_sysvery}. Moreover, if $V^{(k+1)}(x)$  is the positive semi-definite function that satisfies $H(x,V^{(k+1)},\hat{u}^{(k+1)})=0$ with $V^{(k+1)}(0)=0$,  it holds that $V^{*}(x) \leq V^{(k+1)}(x) \leq V^{(k)}(x)$. \label{thorem2}
	\end{theorem}
	\begin{IEEEproof}
		We first prove  that $\hat{u}^{(k+1)} \in \mathcal{A}(\Omega)$. Taking the
		derivative of ${V}^{(k)}(x)$ along the trajectory of system
		$\dot{x}=F(x)+G(x)\hat{u}^{(k+1)}$, it yields
		\begin{align}
		\dot{V}^{(k)}(x)=(V_x^{(k)})^\top F+(V_x^{(k)})^\top G\hat{u}^{(k+1)}. \label{eq:thm2eq1}
		\end{align}
		Since   $\hat{H}(x, {V}^{(k)},\hat{u}^{(k)})=0$, we get
		\begin{equation}
		(V_x^{(k)})^\top F=-(V_x^{(k)})^\top G\hat{u}^{(k)}-Q(x_1)-\hat{U}_n(\hat{u}^{(k)},x).  \label{eq:thm2eq2}
		\end{equation}
		By substituting the term $(V_x^{(k)})^\top F$ with \eqref{eq:thm2eq2}, \eqref{eq:thm2eq1} becomes
		\begin{equation}
		\dot{V}^{(k)}(x)=-Q(x_1)-\hat{U}_n(\hat{u}^{(k+1)},x)-M_t(x), \label{VkX} 
		\end{equation} 
		where $M_t(x)$ is 
		\begin{align}
		M_t(x)=&(V_x^{(k)})^\top G(\hat{u}^{(k)}-\hat{u}^{(k+1)})+\hat{U}_n(\hat{u}^{(k)},x) \notag \\
		&-\hat{U}_n(\hat{u}^{(k+1)},x). \notag 
		\end{align}
		It can be deduced from \eqref{eq:pL} that
		\begin{align}
		(V_x^{(k)})^\top G= -2\hat{\lambda } R \tanh^{-1}(\hat{\lambda}^{-1}\hat{u}^{k+1}). \label{eqVkG}
		\end{align}
		Combined with \eqref{eqVkG} and
		\begin{align}
		&\hat{U}_n(\hat{u}^{k},x)-\hat{U}_n(\hat{u}^{k+1},x) \notag \\
		&=2 \sum_{i=1}^{m}\int_{\hat{u}_i^{(k+1)}}^{\hat{u}_i^{(k)}}r_i\hat{\lambda}_i \left(\tanh^{-1}(s/\hat{\lambda}_i)\right) \mathrm{d} s, \notag 
		\end{align}
		the term $M_t(x)$ can be rewritten as 
		\begin{align}
		M_t(x)=2\sum_{i=1}^{m}r_i\hat{\lambda}_i m^t_i, \notag
		\end{align}
		where $m^t_i$ is 
		\begin{align}
		m^t_i=&\int_{\hat{u}_i^{(k+1)}}^{\hat{u}_i^{(k)}}  \left(\tanh^{-1}(s/\hat{\lambda}_i)\right) \mathrm{d} s \notag \\
		&-\tanh^{-1}(\hat{u}_i^{(k+1)}/\hat{\lambda}_i)(\hat{u}_i^{(k)}-\hat{u}_i^{(k+1)}). \notag 
		\end{align}
		Let the function $\beta(s)$ in \eqref{eq:falpha} be $\beta(s)=\tanh^{-1}(s/\hat{\lambda}_i)$ and we have
		\begin{equation}
		m^t_i=f_\alpha(\hat{u}_i^{(k+1)},\hat{u}_i^{(k)})\ge 0, \notag
		\end{equation}
		which further implies  $M_t(x)\ge 0$. 
		
		Since ${V}^{(k)}(x)\ge 0$ and ${V}^{(k)}(x)= 0$ if and only if $\lVert x_1\rVert =0$, the function ${V}^{(k)}(x)$ can be treated as a Lyapunov function for $x_1$. Then from  \eqref{VkX}, we obtain that $\dot{V}^{(k)}(x)\le 0$ as the functions $Q(x_1)$, $\hat{U}_n(\hat{u}^{(k+1)})$ and $M_t(x)$ are all positive semi-definite. Hence, the $x_1$-system can be stabilized by  the control policy $\hat{u}^{(k+1)}$. Besides, from \eqref{eq:pL}, it can be observed that $\hat{u}^{(k+1)}=0$ if $x_1=0$. Thus, $\hat{u}^{(k+1)}$ is   admissible.
		
		Next, we will prove that $V^{*}(x) \leq V^{(k+1)}(x) \leq V^{(k)}(x)$. As both $\hat{u}^{(k)}$ and   $\hat{u}^{(k+1)}$ are admissible control policies, we have ${V}^{(k)}(x(t=\infty))=0$ and ${V}^{(k+1)}(x(t=\infty))=0$. Taking the
		derivative of ${V}^{(k)}(x)$ and ${V}^{(k+1)}(x)$, respectively, along the trajectory of system
		$\dot{x}=F(x)+G(x)\hat{u}^{(k+1)}$, it yields
		\begin{align}
		&{V}^{(k+1)}(x(t))-{V}^{(k)}(x(t)) \notag \\
		&=-\int_{t}^{\infty}\frac{\mathrm{d}({V}^{(k+1)}(x)-{V}^{(k)}(x))}{\mathrm{d}x}(F+G\hat{u}^{(k+1)})\mathrm{d}\tau \label{dVk}
		\end{align}
		and 
		\begin{align}
		(V_x^{(k+1)})^\top F=&-(V_x^{(k+1)})^\top G\hat{u}^{(k+1)}-Q(x_1) \notag \\
		&  -\hat{U}_n(\hat{u}^{(k+1)}). \label{Vk_1F}
		\end{align}
		By substituting \eqref{eq:thm2eq2} and \eqref{Vk_1F} into \eqref{dVk}, we derive that
		\begin{align}
		&{V}^{(k+1)}(x(t))-{V}^{(k)}(x(t))\notag \\
		=&-\int_t^\infty \left[ (V_x^{(k)})^\top G(\hat{u}^{(k)}-\hat{u}^{(k+1)}) \right.\notag \\
		&\left.+\hat{U}_n(\hat{u}^{(k)})-\hat{U}_n(\hat{u}^{(k+1)}) \right]\mathrm{d}\tau\notag\\
		=&-\int_t^\infty M_t(x(\tau))\mathrm{d}\tau \le 0. \notag
		\end{align}
		Thus ${V}^{(k+1)}(x)\le{V}^{(k)}(x)$ for $\forall x\in\Omega$. Furthermore,  by using the contradiction method, it holds that  $V^{*}(x) \leq V^{(k+1)}(x) \leq V^{(k)}(x)$.
	\end{IEEEproof}

	Theorem~\ref{thorem2} guarantees that the trained control policy is always   admissible  during the process of policy iteration. Meanwhile, the updated control policy is always better than its previous one. Then	to successively solve \eqref{eq:IRLBell2} and  \eqref{eq:pL}, the value function $V(x)$ is approximated by a single-layer neural network, which is
		\begin{equation}
		V(x)=\sum_{j=1}^{M} w_{j}^*\sigma_{j}(x)+\xi(x)=(\boldsymbol{w}^*)^\top  \boldsymbol{\sigma}_{M}(x)+\xi(x), \label{idealeqVM}
		\end{equation}	
		where $\sigma_{j}(x)$ is the activation function and satisfies $\sigma_{j}(x_1=0)=0$; $\xi(x)$ is the  approximation residual error; $w_{j}^*$ represents the ideal weight of the   $j$-th neuron which minimizes the residual error $\xi(x)$; the vector $\boldsymbol{\sigma}_M(x)$ denotes the vector of activation functions   and $\boldsymbol{w}^*$ denotes the ideal weight vector. 
		
		\begin{remark}
			It has been pointed out in \cite{abu2005nearly} that the approximation residual error $\xi(x)$ will converge to zero  when the number of neurons $M\to \infty$. Meanwhile, for fixed $M$,  the approximation residual error $\xi(x)$ is also bounded \cite{hornik1990universal}. In practical implementation, the approximation  residual error is usually reduced by setting   the number of neurons as large as possible. But how to eliminate the approximation residual error $\xi(x)$ completely with a limited   number of neurons still requires further investigation.
		\end{remark}
		
		In order to seek the ideal weight vector $\boldsymbol{w}^*$, the value function $V^{(k)}(x)$ in the $k$-th iteration is approximated  as

	% Let $u^{(0)}$ be an admissible policy, then the iteration between
	
	%1. (policy evaluation) solve for $V^{(i+1)}(\xi(t))$ using the Hamiltonian function \eqref{eq:Halmiton} 
	%\begin{equation}
	%H(V^{(i+1)},u_\theta^*)=0, \label{eq:V_iteration}
	%\end{equation}
	%with the boundary condition $V^{i+1}(0)=0$, and
	%
	%2. (policy improvement) update the control policy using
	%\begin{equation}
	%u_\theta^*(\xi)=\omega_{max}\phi\big(\frac{1}{2\omega_{max}} R^{-1} [0 \quad b_2] V_{\xi}^{*}+c/b_2\big)
	%\end{equation}
	
	%converges to the optimal control policy $u_\theta^*$.
	\begin{figure}[t]
		\centering
		\includegraphics[width=0.48\textwidth,height=3.7 in]{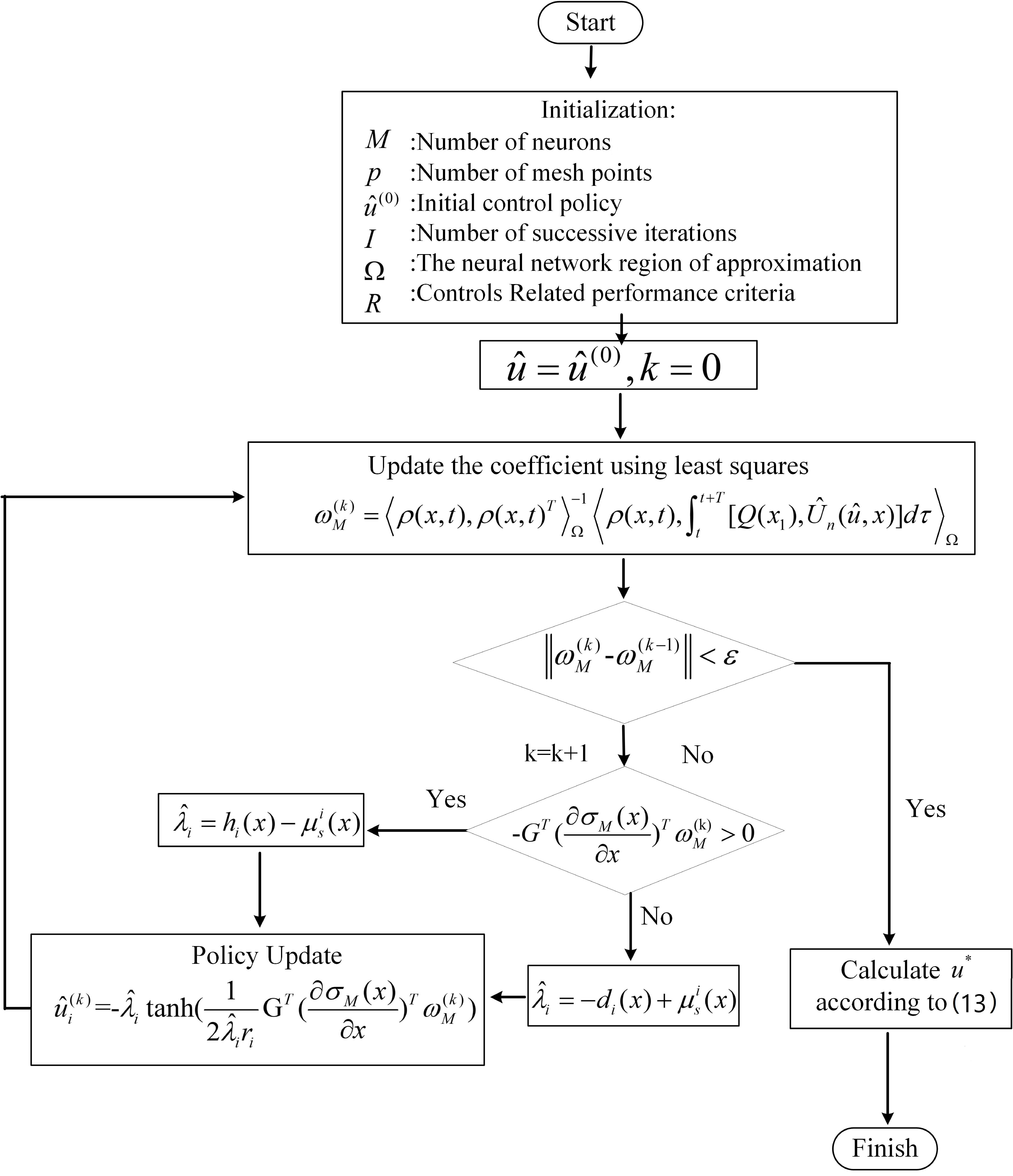}  
		\caption{Flowchart of the proposed IRL algorithm. }  
		\label{fig:flowchart} 
	\end{figure}
	
	\begin{equation}
	V_M^{(k)}(x)=\sum_{j=1}^{M} w_{j}^{(k)}\sigma_{j}(x)=(\boldsymbol{w}_M^{(k)})^\top  \boldsymbol{\sigma}_{M}(x), \label{eqVM}
	\end{equation}
	where  $w_{j}^{(k)}$ and $\boldsymbol{w}_M^{(k)}$ denote the weight of the   $j$-th neuron and the weight vector in the $k$-th iteration, respectively.

	By replacing $V^{(k)}(x)$ in \eqref{eq:IRLBell2} with $V_M^{(k)}(x)$, we have 
	\begin{equation}
	\begin{aligned}
	\left(\boldsymbol{w}_M^{(k)}\right)^{\top} \boldsymbol{\sigma}_{M}&(x(t))= \int_{t}^{t+T}  \big[Q(x_1)+\hat{U}_n(\hat{u},x))\big] \mathrm{d} \tau \\
	&+\left(\boldsymbol{w}_M^{(k)}\right)^{\top}   \boldsymbol{\sigma}_{M}(x(t+T))-e_{M}^{(k)},
	\end{aligned}
	\end{equation}
	where  $ e_{M}^{(k)}$ is the residual error defined by
	\begin{equation}
	\begin{aligned}
	e_{M}^{(k)}=& \int_{t}^{t+T} \big[Q(x_1)+\hat{U}_n(\hat{u},x)\big]  \mathrm{d} \tau \\
	&+\left(\boldsymbol{\omega}_{M}^{(k)}\right)^{\top}  \left[ \boldsymbol{\sigma}_{M}(x(t+T))-\boldsymbol{\sigma}_{M}(x(t))\right]. \label{eq:emi}
	\end{aligned}
	\end{equation}
	Obviously,  the parameter $\boldsymbol{w}_{M}$ should be  tuned  to reduce  the residual error. Here we define a to-be-minimized index  as
	\begin{equation}
	S=\int_{\Omega} |e_{M}^{(k)}(x)|^2 \mathrm{d} x. \notag
	\end{equation}
	To minimize $S$, the weights  $\boldsymbol{w}_M$ is  determined by 
	\begin{equation}
	\left\langle\frac{\mathrm{d} e_{M}^{(k)}(x)}{\mathrm{d} \boldsymbol{w}_M}, e_{M}^{(k)}(x)\right\rangle_\Omega= 0,  \label{eq:integ_error}
	\end{equation}
	where the notation $\langle f(x), g(x)\rangle=\int_{\Omega} f(x) g(x)^\top  \mathrm{d} x$ denotes the Lebesgue integral. Let $ \varrho(x,t)= \boldsymbol{\sigma}_{M}(x(t+T))-\boldsymbol{\sigma}_{M}(x(t)) $. By substituting \eqref{eq:emi} into \eqref{eq:integ_error}, one has
	\begin{equation}\begin{array}{l}
	\left\langle\varrho(x,t),  \int_{t}^{t+T} \big[Q(x_1)+\hat{U}_n(\hat{u},x)\big]  \mathrm{d} \tau  \right\rangle_{\Omega}\\
	+\left\langle\varrho(x,t),\varrho(x,t)^\top \right\rangle_{\Omega}\boldsymbol{w}_M^ {(k)} =0. \label{eq51}
	\end{array}\end{equation}

		To solve $\boldsymbol{w}_M^ {(k)}$, we impose the following assumption in the spirit of persistent excitation (PE) condition.
		\begin{assump}
			For all admissible control policies $\mu(x)\in \mathcal{A}(\Omega)$, there exist constants $\overline{m}_c>0$ and  $\gamma_c>0$ such that 
			\begin{equation}
			\frac{1}{m_c}\sum_{i=1}^{m_c}\varrho(x,t_i)\varrho(x,t_i)^\top \ge \gamma_c I_M
			\end{equation} 
			for all   $m_c\ge \overline{m}_c$. \label{assump:PE}
		\end{assump}
		
		Let $p$ represent the number of points in the sample set $\Omega$. If Assumption~2 holds and $p\ge\overline{m}_c$, it can be inferred that $\left\langle\varrho(x,t),\varrho(x,t)^\top \right\rangle_{\Omega}$ is invertible.
	Thus,  based on \eqref{eq51}, $\boldsymbol{\omega}_M^{(k)}$ is updated as
	\begin{align}
	\boldsymbol{w}_M^{(k)} =-&\left\langle\varrho(x,t), \varrho(x,t)^\top \right\rangle^{-1}_\Omega \times \notag\\
	&\left\langle\varrho(x,t),  \int_{t}^{t+T} \big[Q(x_1)+\hat{U}_n(\hat{u},x)\big]  \mathrm{d} \tau  \right\rangle_{\Omega}. \label{eq:update} 
	\end{align}
	
	%\begin{figure}
	%	\centering  
	%	\includegraphics[width=0.4\textwidth,height=4in]{picture/flowchart}  
	%	\caption{The flowchart of the algorithm for nearly optimal saturated neurocontrol. }  
	%	\label{fig:flowchart}  
	%\end{figure}
	
	Afterwards, in order to solve \eqref{eq:update}, an iterative algorithm proposed by  \cite{abu2005nearly} is adopted. Given some points over the integration region on $\Omega$,  define
	\begin{align}
	L&=\big[\varrho(x,t)|_{x_1},\cdots,\varrho(x,t)|_{x_p} \big], \notag \\
	Y&=\bigg[\int_{t}^{t+T} (Q(x_1)+ \hat{U}_n(\hat{u},x))\mathrm{d} \tau|_{x_1},\cdots,   \notag \\
	&\int_{t}^{t+T} (Q(x_1)+ \hat{U}_n(\hat{u},x) )\mathrm{d} \tau|_{x_p}\bigg]. \notag 
	\end{align}
	Then we get 
	\begin{align}
	&\left\langle\varrho(x,t), \varrho(x,t)^\top \right\rangle_\Omega=\lim\limits_{\lVert \delta x \rVert \to 0}(L^\top L)\delta x \notag \\
	&\left\langle\varrho(x,t), \int_{t}^{t+T} \big[Q(x_1)+ \hat{U}_n(\hat{u},x) \big]\mathrm{d} \tau \right\rangle_\Omega=\lim\limits_{\lVert \delta x \rVert \to 0}(L^\top Y)\delta x.  \label{eqLTY}
	\end{align}
	By using \eqref{eqLTY}, we rewrite \eqref{eq:update}  as
	\begin{equation}
	\boldsymbol{w}_M^{(k)}=-(L^\top L)^{-1}(L^\top Y). \label{wmupdate}
	\end{equation}

		\begin{remark}
			Assumption~\ref{assump:PE} is   a common assumption in the current study of ADP \cite{bian2014adaptive,modares2014optimal,Li2017Off}. Theoretically, the validity of Assumption~\ref{assump:PE} is related to the integration time $T$ and the richness of the collected samples. It has been pointed out in \cite{vrabie2009neural} that if a proper  integral  time $T$ is selected, the sample size $p$ just needs to  be no smaller  than $M$ such that the  matrix $\left\langle\varrho(x,t),\varrho(x,t)^\top \right\rangle_{\Omega}$ is invertible. However, it still remains an unsolved problem on how to select such a proper  integral  time $T$.  As a consequence, in practical implementation,   to enrich the diversity of samples so that  Assumption~\ref{assump:PE} holds,    the control input in the training phase is expected to be persistently exciting, which is usually guaranteed by adding a small exploration noise to the original control input  \cite{vrabie2009neural}. After the training is finished, the exploration noise is removed.  Meanwhile,  the size of the sample set $\Omega$ is usually chosen as large as possible  to guarantee  $p\ge m_c$. 
		\end{remark}

   According to the definition of $\hat{\lambda}_i$ in \eqref{eq:overlamda},   the sign of  each element $\hat{\mu}_i(x)$  of $\hat{\mu}(x)$ should be evaluated in advance. From the structure of $\hat{\mu}_i(x)$ in \eqref{eq:pL}, it can be found that the sign of $\hat{u}_i(t)$ is the same as that of the $i$-th element of $-G^{\top}  (x) V_x^{(k)}(x)$.  Thus,  the matrix $\hat{\lambda}$ can be determined by calculating $-G^{\top}  (x) V_x^{(k)}(x)$ in advance. Specifically, the $i$-th element of $\hat{\lambda}$ on the diagonal is determined by 
	\begin{equation}
	\hat{\lambda}_i=\left\{\begin{array}{lll}
	h_i(x)-\mu_s^i(x),    \quad \   &\text{if} \quad &z_i \ge 0, \\
	- d_i(x)+\mu_s^i(x),   \quad   &\text{if} \quad &z_i < 0,
	\end{array}\right. \notag
	\end{equation}
	where $z_i$ is the  $i$-th element of  $-G^{\top}  (x ) V_x^{(k)}(x)$.
	
	The iterations is terminated when the error of the coefficients obtained at two consecutive steps is smaller than a given threshold $\epsilon$. The flow chart of the proposed IRL algorithm is presented in Fig.~\ref{fig:flowchart}.
	\begin{remark}
	As the traditional admissible control policies can stabilize the system, the initial control policy $u(t)$ defined by \eqref{eq_uus}  with the NN weight being zero vector is also admissible. 	Thus  in the PI algorithm described above,  the NN weight can be initialized as zero vector directly.
	\end{remark}

	\section{Application  to the optimal UAV circumnavigation problem} \label{sec:5}
	
	In this section,  we employ the method proposed in this paper to solve a  practical problem: the optimal UAV circumnavigation control problem. As the target's position  is usually estimated from the on-board sensor measurement, which often contains noise, a filter  like Extended Kalman Filter (EKF) is needed. It has been shown in \cite{ponda2009trajectory} that the performance of the filter is dependent on the UAV's trajectory. In this section, we intend to design an optimal circumnavigation controller based on the Fisher information,  which  can quantify the information provided by the sensor measurement  \cite{Wang2011Bearings}. Generally speaking, the more Fisher information the UAV gains,   more accurate the estimated target position will be  \cite{chen2016cooperative}.  Specifically, the UAV is controlled to circumnavigate around the target  (see Fig.~\ref{fig:situation}), while minimizing an objective function involving Fisher information in the circumnavigation trajectory.  A simulation result is presented in which the performance of the control law designed by our method is compared with the method  proposed in  \cite{dong2019flight}. 
	\subsection{Problem formulation of the optimal UAV circumnavigation }
	
	Consider a fixed-wing UAV, whose  kinematic model is described by
	\begin{equation}
	\left\{\begin{array}{l}{\dot{x}_p=v \cos \theta}, \\ 
	{\dot{y}_p=v \sin \theta}, \\
	{\dot{\theta}=u_\theta}, \\
	v=u_v, \end{array}\right.  \label{eq:UAV_kinematic}
	\end{equation}
	where $(x_p,y_p)$ is the position of the UAV and $\theta$ denotes the  heading angle of the UAV;   $v$  is the UAV's linear velocity; $u_v$ and $u_\theta$ are the control inputs.  The height of the UAV  is assumed to be held constant. Owing to the  roll angle constraint of the fixed-wing UAV, the following unsymmetrical input constraint is enforced on the UAV:
	
		\begin{equation}
 	-\frac{\omega_{\min} }{1+0.02v}\leq u_\theta \leq \frac{\omega_{\max}}{1+0.02v}, \notag 
		\end{equation}
	where the constants $\omega_{\max},\omega_{\min}>0$ and $\omega_{\max}\neq \omega_{\min}$. It is obvious that the input saturation constraint is unsymmetrical and depends on the UAV's linear velocity.  Note that the linear velocity $v$ is not constant but is  dependent on the UAV's state, which will be illustrated later.

	\begin{figure}[t]
		\centering
		\includegraphics[width=0.45\textwidth,height=1.6 in]{picture/situation} 
		\caption{The illustration of circumnavigation around a ground moving vehicle.}  
		\label{fig:situation}   
	\end{figure}

 Let  $s_t=(x_t,y_t)^\top \in \mathbb{R}^2$ represent  	the position of the moving target. It is assumed that the ground target moves with a constant linear velocity $v_t$ and the dynamics of the target is described by
	\begin{equation}
	\left\{\begin{array}{l}
	\dot{x}_t=v_t \cos\theta_t, \\ 
	\dot{y}_t=v_t \sin \theta_t, \\
	\dot{\theta}_t=h(\theta_t),
	\end{array}\right. \label{eq:car_kinematic}
	\end{equation}
	where $\theta_t$ is the heading of the target, and $h(\theta_t)$ is an unknown function.

	As the UAV is expected to hold a constant angular speed with a preset circumnavigation radius, the relative speed $v_r$ of the UAV is also expected to be constant. The  relative angle   of the UAV  w.r.t. the target  is denoted by $\theta_r$, which  satisfies 
	\begin{equation}
	\left\{\begin{array}{l}
	v_r \cos \theta_r=v\cos\theta-v_t\cos\theta_t, \\ 
	v_r \sin \theta_r=v\sin\theta-v_t\sin\theta_t.
	\end{array}\right. \label{eq:v_r}
	\end{equation}
The linear speed of the UAV is dependent on the UAV's state and  can be obtained from   \eqref{eq:v_r} as
\begin{equation}
v=v_t\cos(\theta-\theta_t)+\sqrt{v_t^2\cos^2(\theta-\theta_t)+v_r^2-v_t^2}. \label{linerv}
\end{equation}

	The UAV is assumed to utilize a radar as the measurement  sensor. Let $s_r=(x_r,y_r)^\top =(x_p-x_t,y_p-y_t)^\top $ denote the relative position between the UAV and the target. Then with the aid of the radar, the UAV can sense the range and bearing information with the observation  model described by 
	\begin{equation}
	\zeta(t)=\mathcal{Z}(s_r)+\chi(t), \notag
	\end{equation}
	where $\zeta(t)$ is the  sensor measurement; $\chi(t)$ is the measurement noise;  $\mathcal{Z}(s_r)$ is the observation function  defined by
	\begin{equation}
	\mathcal{Z}(s_r)= \left[\begin{array}{c}{r} \\ {\varphi}\end{array}\right]=
	\left[\begin{array}{c}{\sqrt{x_{r}^{2}+y_{r}^{2}+h^{2}}} \\ {\arctan \left(y_{r} / x_{r}\right)}\end{array}\right], \label{eq:observer_mdl}
	\end{equation}
	where  $h$  is the height of the UAV;    $r$    is  the  distance   between the UAV and the target in 3-dimensional space; $\varphi$  is the bearing  angle between the target  and the UAV in the plane.

	The   circumnavigation radius  error $e_r$  is defined by
	\begin{equation}
	e_r=r_h-r_d,
	\end{equation}
	and  a state $\eta$ is defined by
	\begin{equation}
	\eta=\frac{\pi}{2}-(\theta_r-\varphi), \label{eq:eta_def}
	\end{equation}
	where $r_d$ is the desired circumnavigation radius around the target and $r_h=\sqrt{x_r^2+y_r^2}$ represents the current circumnavigation radius. Note that the desired circumnavigation  around the target is achieved if the conditions $e_r= 0$ and $\eta= 0$ are satisfied \cite{yu2020optimal}. Thus, to make the state $e_r$ and $\eta$ converge to zero, the dynamics of $e_r$ and $\eta$ will be analyzed first. The dynamics of $e_r$ is 
	\begin{align}
	\dot{e}_r&=\dot{r}_h=\frac{x_r\dot{x}_r+y_r\dot{y}_r}{r_h} \notag =v_r \cos\varphi cos\theta_r+v_rsin\varphi \sin\theta_r \notag \\
	&=v_r\cos(\varphi-\theta_r)=v_r\sin\eta.
	\end{align}
%	Then according to \eqref{eq:v_r}, one has
%	\begin{equation}
%	v^2+v_t^2-2vv_tcos(\theta-\theta_t)=v_r^2. \label{eq:relation_v}
%	\end{equation} 
	Deriving both sides of the equation \eqref{eq:v_r} by time $t$, one has
	\begin{align}
	-v_r\sin\theta_r\dot{\theta}_r=v_t\sin \theta_t\dot{\theta}_t+\dot{v}&\cos\theta-v\sin\theta\dot{\theta}, \notag  \\
	v_r\cos\theta_r\dot{\theta}_r=-v_t\cos  \theta_t\dot{\theta}_t+\dot{v}&\sin\theta+v\cos\theta\dot{\theta}. \notag 
	\end{align}
	By eliminating  $\dot{v}$, it yields
	\begin{align}
	\dot{\theta}_r&=\frac{v}{ v_r\cos(\theta_r-\theta)}\dot{\theta}-\frac{v_t\cos(\theta-\theta_t)}{ v_r\cos(\theta_r-\theta)}\dot{\theta}_t. \label{eq:dotthetar}
	\end{align}
	Further, the dynamics of $\varphi$ can be obtained by
	\begin{align}
	\dot{\varphi}=&\frac{x_r\dot{y}_r-y_r\dot{x}_r}{r_h^2}=\frac{\cos\varphi(v_r\sin\theta_r)-\sin\varphi(v_r\cos\theta_r)}{r_h} \notag \\
	=&\frac{v_r\sin(\theta_r-\varphi) }{r_h}=\frac{v_r\cos\eta }{r_d+e_r}.\label{eq:dotphi}
	\end{align}
	By combining \eqref{eq:dotthetar} and \eqref{eq:dotphi}, it yields
	\begin{align}
	\dot{\eta}=&\dot{\varphi}-\dot{\theta}_r \notag \\
	=&\frac{v_r\cos\eta }{r_d+e_r}+\frac{v_t\cos(\theta-\theta_t)}{ v_r\cos(\theta_r-\theta)}\dot{\theta}_t-\frac{v}{ v_r\cos(\theta_r-\theta)}\dot{\theta},\notag
	\end{align}
	where $\theta_r$ is determined by \eqref{eq:v_r} and can be calculated by
	\begin{align}
	\theta_r={\rm atan2}(v\sin\theta-v_t\sin\theta_t,v\cos\theta-v_t\cos\theta_t). \notag
	\end{align}
	Define a state variable $x=(e_r,\eta,\theta,\theta_t)^\top $. Then the $x$-dynamics is described by
	\begin{equation}
	\left\{\begin{array}{l}
	\dot{e}_r=v_r\sin\eta, \\ 
	\dot{\eta}=\frac{v_r\cos\eta }{r_d+e_r}+\frac{v_t\cos(\theta-\theta_t)}{ v_r\cos(\theta_r-\theta)}h(\theta_t)-\frac{v}{ v_r\cos(\theta_r-\theta)}u_\theta, \\
	\dot{\theta}=u_\theta, \\
	\dot{\theta}_t=h(\theta_t),
	\end{array}\right. \label{eq:r_dynamic}
	\end{equation}
			where the linear speed   $v$  of the UAV is given by  \eqref{linerv}.

		It can be observed that system \eqref{eq:r_dynamic} is a NF system as $\dot{\eta}\neq 0$ when $e_r$, $\eta$ and $u_\theta$ are all 0. 	By using the method proposed in Section~\ref{sec:design}, the control policy $u_\theta(t)$ is designed as
	\begin{equation}
	u_\theta(t)=u_s(t)+\hat{u}(t), \label{hatu2}
	\end{equation}
	where $u_s(t)$ is the initial  control policy and $\hat{u}(t)$ is the virtual input to be designed. Specifically, we adopt the vector field (VF) method proposed in  \cite{dong2019flight} as the initial admissible control policy $u_s(t)$, which is  described by
	\begin{equation}
	u_s=\left\{\begin{array}{lll}
	-k(\theta-\theta_d),   &\text{if\ }  & k(\theta_d-\theta)\in [-\frac{\omega_{\min} }{1+0.02v},\frac{\omega_{\max} }{1+0.02v}],    \\ 
	\frac{\omega_{\max} }{1+0.02v},  &\text{if\ }  & k(\theta_d-\theta)>\frac{\omega_{\max} }{1+0.02v} , \\
	-\frac{\omega_{\min} }{1+0.02v},   &\text{if\ }  &k(\theta_d-\theta)<-\frac{\omega_{\min} }{1+0.02v},
	\end{array}\right. \notag
	\end{equation}
	where $\theta_d$ is determined by 
	\begin{equation}
	\left[\begin{array}{c}{\cos\theta_d} \\ {\sin\theta_d}\end{array}\right]=\frac{-v_r}{r_{h}\left(r_{h}^{2}+r_d^{2}\right)}\left[\begin{array}{l}{x_{r}\left(r_{h}^{2}-r_d^{2}\right)+y_{r}\left(2 r_d r_{h}\right)} \\ {y_{r}\left(r_{h}^{2}-r_d^{2}\right)-x_{r}\left(2 r_d r_{h}\right)}\end{array}\right]. \notag
	\end{equation}	
	By substituting \eqref{hatu2} into \eqref{eq:r_dynamic}, the $x$-dynamics becomes
	\begin{equation}
	\left\{\begin{array}{l}
	\dot{e}_r=v_r\sin\eta, \\ 
	\dot{\eta}=\frac{v_r\cos\eta }{r_d+e_r}+\frac{v_t\cos(\theta-\theta_t)}{ v_r\cos(\theta_r-\theta)}h(\theta_t)-\Lambda u_s-\Lambda\hat{u}, \\
	\dot{\theta}=u_s+\hat{u},\\
	\dot{\theta}_t=h(\theta_t).
	\end{array}\right. \label{eq:r_dynamic2}
	\end{equation}
	where $\Lambda=v/(v_r\cos(\theta_r-\theta))$ 	and the virtual input $\hat{u}$ is constrained by
	\begin{equation}
	u_s-\frac{\omega_{\min} }{1+0.02v} \le \hat{u} \le u_s+\frac{\omega_{\max} }{1+0.02v}. \notag
	\end{equation}

	%\begin{figure}[t]
	%	\centering
	%	\includegraphics[width=0.45\textwidth,height=2.2 in]{picture/Q}  
	%	\caption{The illustration of $\hat{Q}$ gained in per simulation step. }  
	%	\label{fig:Q} 
	%\end{figure}
	
	Up to now,  the system dynamics for the UAV circumnavigation problem has been formulated. Then a to-be-minimized optimization criterion is  needed. Firstly, using the method proposed in Section~\ref{sec:design}, the control cost function  $\hat{U}_n(\hat{u},x)$  is   defined as
	\begin{equation}
	\hat{U}_n(\hat{u},x)=2 \int_{0}^{\hat{u}}\hat{\lambda} \tanh^{-1}(s/\hat{\lambda})  r  \mathrm{d} s,  \notag
	\end{equation}
	where the constant $r>0$ and  $\hat{\lambda}\in\mathbb{R}$ is defined by
	\begin{equation}
	\hat{\lambda}=\left\{\begin{array}{lll}
	u_s(t)+\dfrac{\omega_{\max} }{1+0.02v},    \quad   &\text{if} \quad &\hat{u}\ge 0, \\ 
	u_s(t)-\dfrac{\omega_{\min} }{1+0.02v},   \quad   &\text{if} \quad &\hat{u} < 0.
	\end{array}\right.  \notag
	\end{equation}

	Next, a function representing the state cost will be constructed based on the so-called   accumulative   information \cite{ponda2009trajectory}. To quantify the utilization of the sensor data,  we set up the  optimization criterion by exploiting the accumulative   information $\mathcal{D}$  based on the Fisher  information metric, which is    
	\begin{align}
	\mathcal{D}=\int_{t_0}^{\infty}\sqrt{L(r_h,\eta)} \mathrm{d}t, \label{eq:D}
	\end{align} 
	where $L(r_h,\eta)$ is
	\begin{equation}
	L(r_h,\eta)=\frac{v_r^2r_h^2}{r^6\sigma_r^2}\sin^2\eta+\frac{8r_h^2v_r^2}{r^4}\sin^2\eta+\frac{v_r^2}{r_h^2\sigma_\varphi^2}\cos^2\eta;  \label{Lrh2}
	\end{equation}
 $\sigma_r$  and  $\sigma_\varphi$  are two constants representing the standard deviations of the rang and bearing measurements, respectively. The detailed   derivation for the formula \eqref{Lrh2} can be found in our previous work \cite{yu2020optimal}.

	It can be deduced intuitively from \eqref{eq:D} and \eqref{Lrh2} that the UAV will obtain more accumulative   information  in the unit time if the circumnavigation radius $r_h$ decreases. In other words, the UAV  will fly just above the ground target  if one directly takes    \eqref{eq:D} as a  to-be-maximized performance index. However,  the UAV is expected to circumnavigate around the target with a given radius, which implies the  performance index should reach its extremum at $r_h=r_d$. To achieve this,   similar to our previous work \cite{yu2020optimal}, a small variation is made based  on the definition of the accumulative   information  and a to-be-minimized performance index is defined as 
	\begin{align}
	&\mathcal{J}=\int_{0}^{\infty}\left[ (Q_{\max}-\hat{Q}(r_h,\eta) )/Q_{\max}+\hat{U}_n(\hat{u},x)\right]\mathrm{d}\tau, \label{eq:performance_index} \\
	&\hat{Q}(r_h,\eta)= \sqrt{L(r_h,\eta) }\tanh(r_h-\kappa),  \notag  
	\end{align}
	where   $\hat{Q}(r_h,\eta)$ is a function which varies slightly from the function  $\sqrt{L(r_h,\eta)}$; the constant $Q_{\max}= \hat{Q}(r_d,0)$ equals the value of $\hat{Q}(r_h,\eta)$ when $r_h=r_d$ and $\eta=0$;   $\kappa$ is a bias constant determined by
	\begin{equation}
	\frac{\mathrm{d} \hat{Q}(r_h,\eta)}{d r_h}(r_h=r_d,\eta=0)=0.  \label{eq:DL_Dr}
	\end{equation}
Specifically,	the exact form of \eqref{eq:DL_Dr} can be obtained by
	\begin{align}
	\alpha(\kappa)=-\tanh(r_d-\kappa)+r_d(1-\tanh^2(r_d-\kappa))=0. \notag
	\end{align}
	Note that the function $\alpha(\kappa)$ is a monotonically decreasing function and $\kappa<r_d$. Thus the constant $\kappa$ can be calculated by using the numerical stepwise  methods.  Affected by the term $\tanh(r_h-\kappa)$, the function $\hat{Q}(r_h,\eta)$ reaches its maximum at $r_h=r_d$ and $\eta=0$. If the UAV is controlled by the optimal control policy $u_\theta^*$ w.r.t. the performance index \eqref{eq:performance_index}, the desired circumnavigation will be achieved while the accumulative information is maximized~\cite{yu2020optimal}.

	Given   system \eqref{eq:r_dynamic2} and the performance index \eqref{eq:performance_index}, the optimal virtual input $\hat{u}^*(t)$ can be obtained through the methods proposed in this paper. Further the optimal control policy $u_\theta^*(t)$ is obtained by $	u_\theta^*(t)=u_s(t)+\hat{u}^*(t)$.
%	\begin{equation}
%	u_\theta^*(t)=u_s(t)+\hat{u}^*(t). \notag
%	\end{equation} 

	\subsection{Simulation result and the comparison with the existing control law}
\begin{figure}[t]
		\centering
		\includegraphics[width=0.35\textwidth,height=1.8 in]{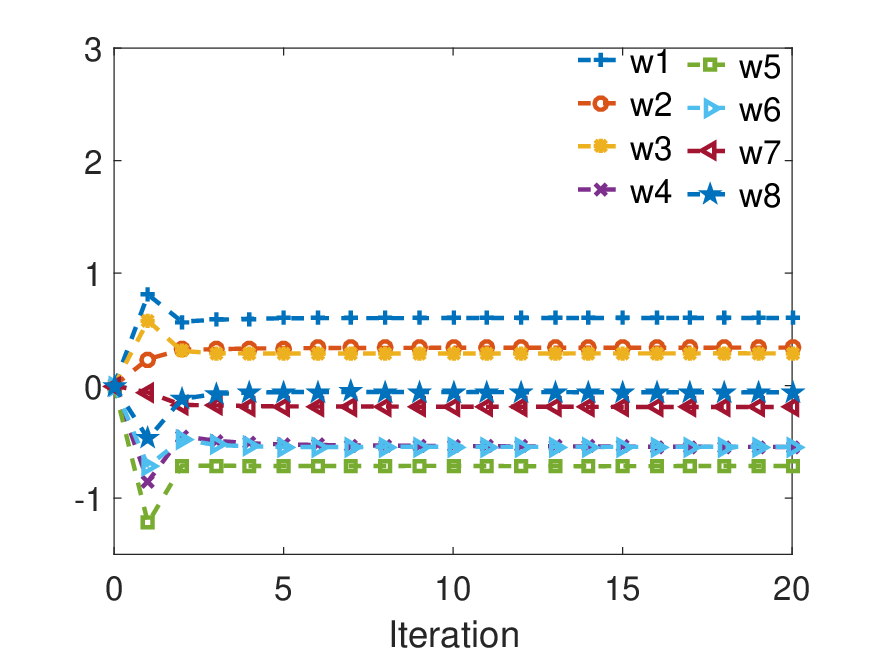}  
		\caption{The  convergence of eight representative NN weights. }  \label{fig:NNweight}  
	\end{figure}

	To validate the performance of the designed circumnavigation  control law,  a numerical simulation is presented in this section. In the simulation, the UAV is expected to circumnavigate around a ground target moving with a linear speed of 5 $m/s$.  The desired circumnavigation radius and the height of the UAV are   set as 50 $m$ and    80 $m$, respectively. The relative linear speed of the UAV w.r.t. the target is 10 $m/s$. The standard deviation parameters $\sigma_r$ and $\sigma_\varphi$  are assumed to be $\sigma_r=2\times 10^{-3}/m$ and $\sigma_\varphi=1.5 \times 10^{-4} \pi$ rad, respectively.  The constants $w_{\max}$ and $w_{\min}$ are set as 1.5 rad/s and 1.2 rad/s, respectively.  The function $h(\theta_t)$ which determines the angular speed of the target is appointed as
	\begin{equation}
	h(\theta_t)=0.5-0.5\sin^2(\theta_t).
	\end{equation}
  In the simulation, the following neural network is utilized to approximate the value function: 
	\begin{equation}
	\begin{aligned} V_{350}\left(e_r,\eta,\theta, \theta_t\right)= \sum_{i=1}^{35}\sum_{j=1}^{10}w_{10(i-1)+j}a_ib_j, \end{aligned} \notag
	\end{equation}
	where $a_i$ and $b_j$ are, respectively, the $i$-th and $j$-th elements of the vectors $\boldsymbol{\vec{a}}$ and $\boldsymbol{\vec{b}}$   defined as follows:
	\begin{align}
	&\boldsymbol{\vec{a}}=[e_r^2,e_r\eta,\eta^2, e_r^4,e_r^3\eta,e_r^2\eta^2,e_r\eta^3,\eta^4, e_r^6, e_r^5\eta, e_r^4\eta^2,e_r^3\eta^3, \notag \\
	& e_r^2\eta^4,e_r^5\eta,\eta^6, e_r^8,e_r^7\eta,e_r^6\eta^2,e_r^5\eta^3,e_r^4\eta^4,e_r^3\eta^5,e_r^2\eta^6,e_r\eta^7, \eta^8,\notag \\
	& e_r^{10},e_r^9\eta,e_r^8\eta^2,e_r^7\eta^3,e_r^6\eta^4,e_r^5\eta^5,e_r^4\eta^6,e_r^3\eta^7, e_r^2\eta^8,e_r\eta^9,\eta^{10} ]^\top,\notag \\
	&\boldsymbol{\vec{b}}=[1,\theta,\theta_t,\theta^2,\theta\theta_t,\theta_t^2,\theta^3,\theta^2\theta_t,\theta\theta_t^2,\theta_t^3]^\top.  \notag
	\end{align}
	The simulation is conducted with a sample frequency of 200Hz.   For each iteration, 40000 samples are collected and used to update the control policy.  The convergence of eight representative weights of the neural network is demonstrated in Fig.~\ref{fig:NNweight}.  It can be observed that after 4 iterations, the weights are  all nearly convergent. Thus, the stableness and convergence of the proposed IRL-based PI algorithm are verified by Fig.~\ref{fig:NNweight}.  	

	\begin{figure*}[t]
		\begin{minipage}[t]{0.5\linewidth}
			\centering
			\includegraphics[width=0.8\textwidth,height=1.8 in]{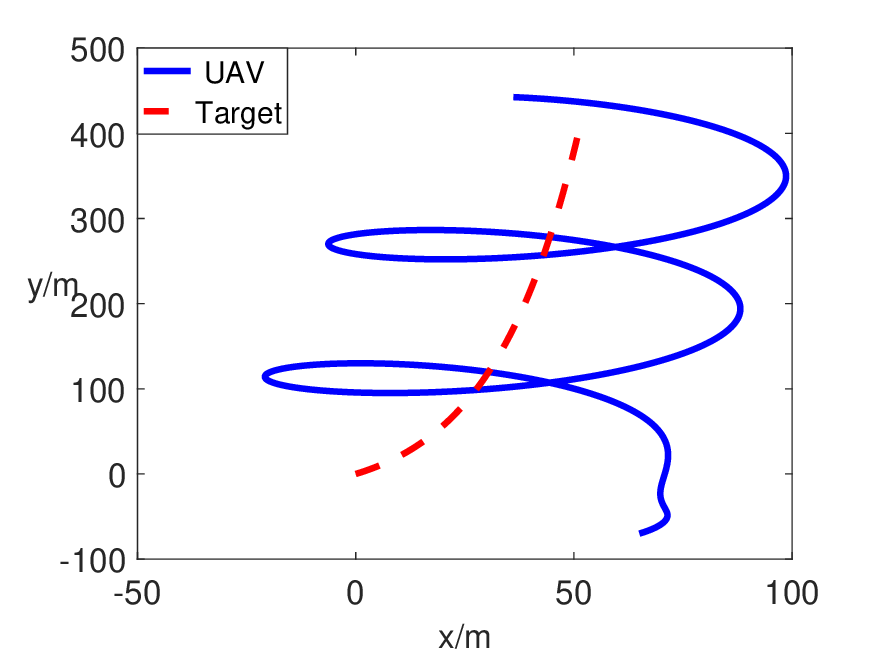}  
			\caption{The illustration of the trajectories of the ground target and the UAV. }  
			\label{fig:trjectory}  
		\end{minipage}%
		\quad
		\begin{minipage}[t]{0.5\linewidth}
			\centering
			\includegraphics[width=0.8\textwidth,height=1.8 in]{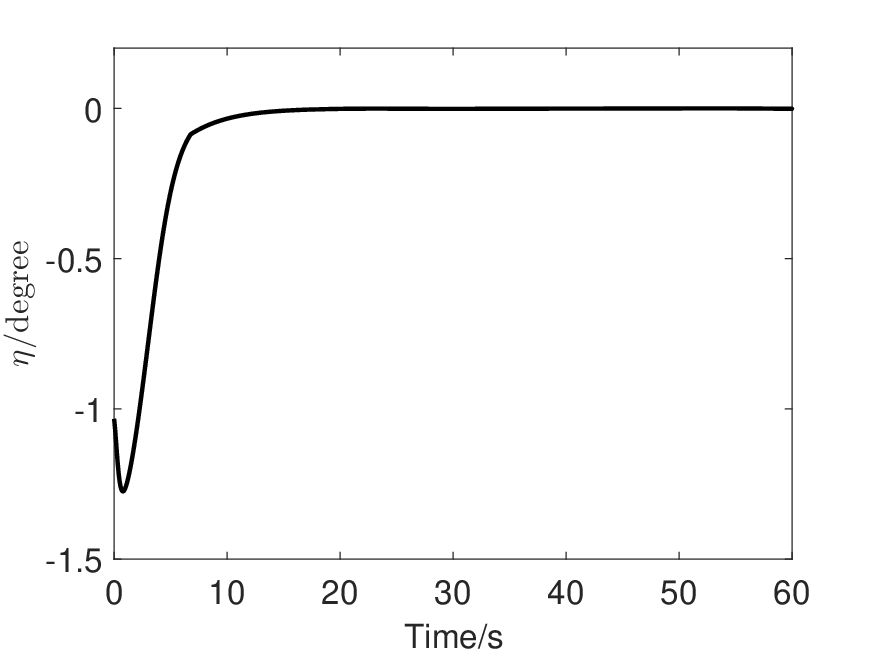}  
			\caption{The variation of $\eta$ during the circumnavigation. }  
			\label{fig:eta} 
		\end{minipage}
		
		\begin{minipage}[t]{0.5\linewidth}
			\centering
			\includegraphics[width=0.8\textwidth,height=1.8 in]{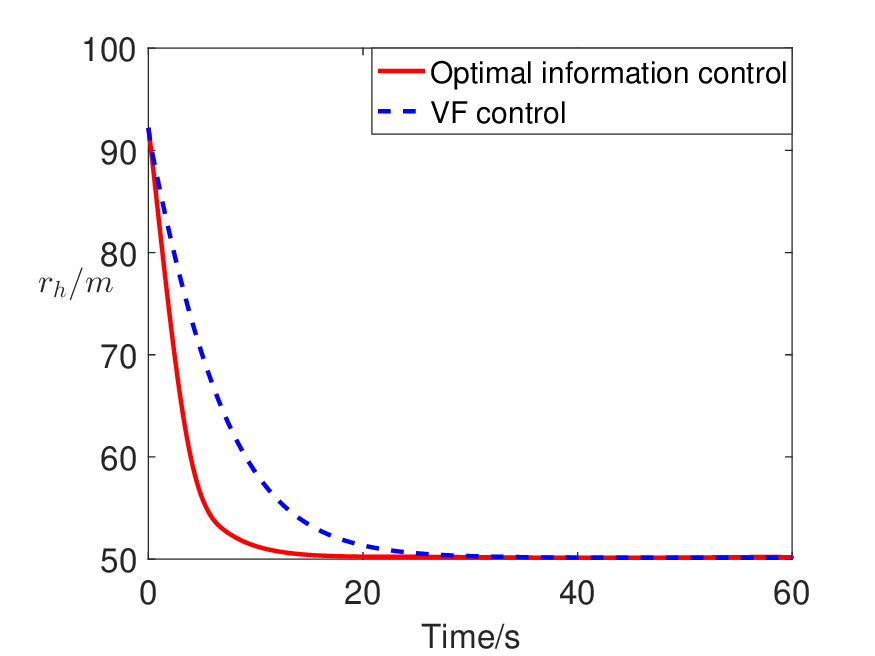}  
			\caption{The variation of the relative distances $r_h$ during the circumnavigation. }  
			\label{fig:dis}  
		\end{minipage}%
		\quad
		\begin{minipage}[t]{0.5\linewidth}
			\centering
			\includegraphics[width=0.8\textwidth,height=1.8 in]{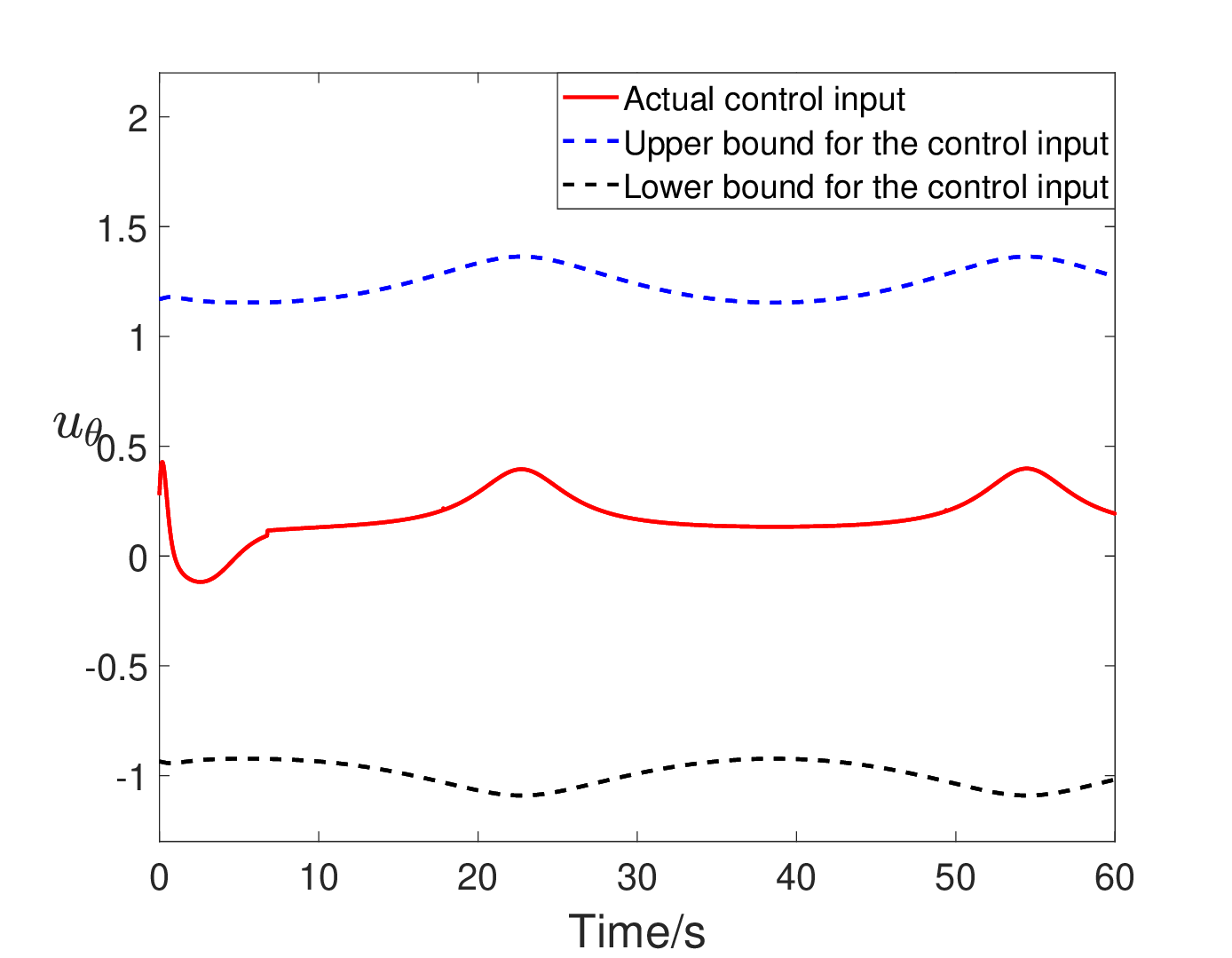}  
			\caption{The variation of the control input. }  
			\label{fig:input} 
		\end{minipage}

		\begin{minipage}[t]{0.5\linewidth}
			\centering
			\includegraphics[width=0.8\textwidth,height=1.8 in]{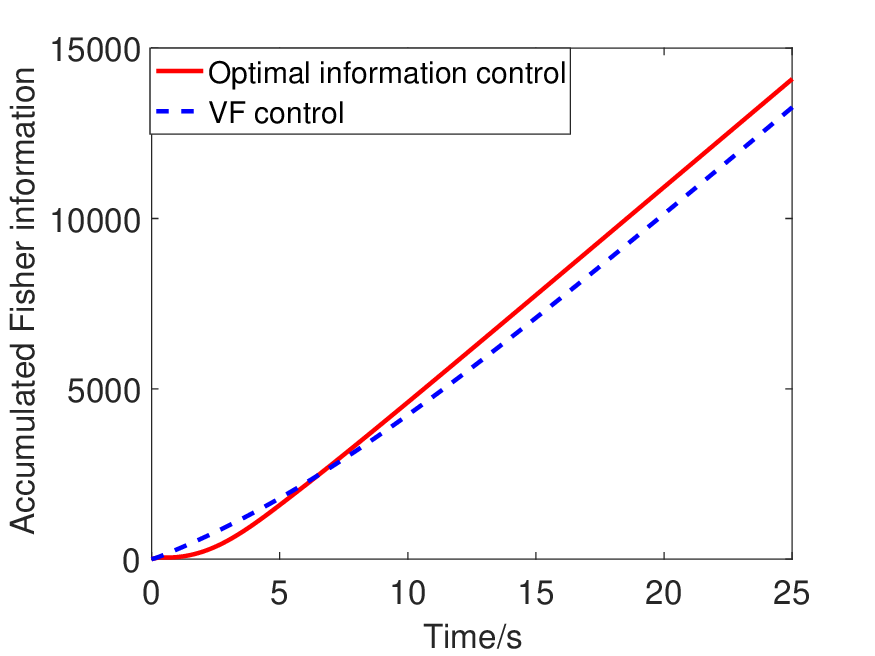}  
			\caption{The variation of the accumulative   information  during the circumnavigation. }  
			\label{fig:Fisher} 
		\end{minipage}%
		\quad
		\begin{minipage}[t]{0.5\linewidth}
			\centering
			\includegraphics[width=0.8\textwidth,height=1.8 in]{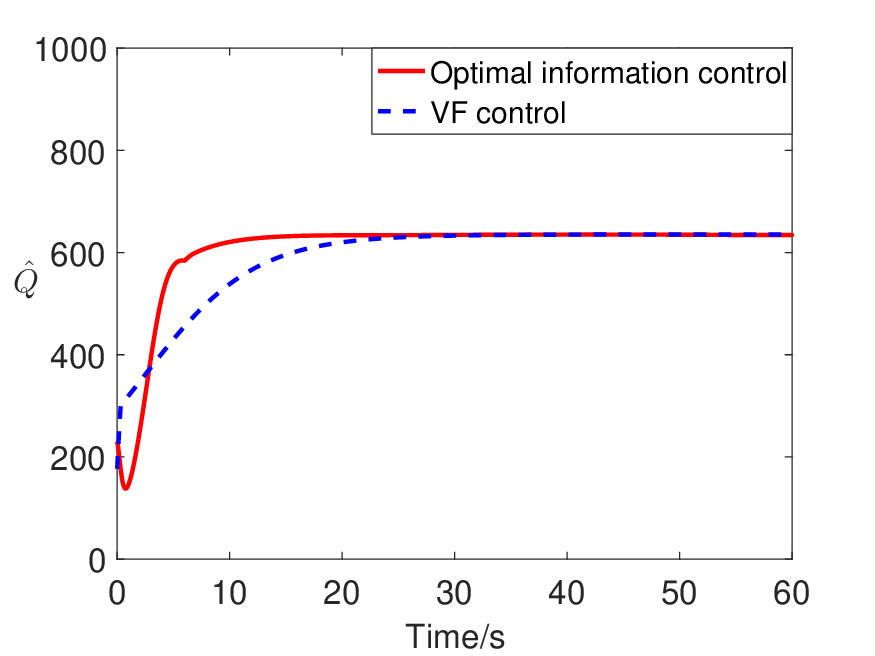}  
			\caption{The variation of $\hat{Q}$ gained  per simulation step. }  
			\label{fig:Q} 
		\end{minipage}
		
	\end{figure*}

	The trajectories of the ground target and  the UAVs controlled by our designed control law are demonstrated in Fig.~\ref{fig:trjectory}. The variation of   the circumnavigation radius $r_h$ controlled by our method is demonstrated in Fig.~\ref{fig:dis}  (red solid line). It is illustrated that  the circumnavigation radius $r_h$  converges to the preset radius 50 $m$.  Fig.~\ref{fig:eta} is the illustration of the variation of the state $\eta$ controlled by the proposed algorithm  during the simulation, which converges to zero at around 17s. The  control input of the UAV   by our method is illustrated in Fig.~\ref{fig:input}. Note that the upper bound (blue dash line in Fig.~\ref{fig:input}) and lower bound (black dash line in Fig.~\ref{fig:input}) for the control input are changing over time as the speed of the UAV varies.   Fig.~\ref{fig:input} shows that the control input of the UAV is always within the allowed range during the process of  the  circumnavigation. 
	
	To demonstrate the validity of our method,  the  proposed control law is further compared with the vector field (VF) guidance law \cite{dong2019flight}.  The comparison of the relative distance  $r_h$ between the two methods is demonstrated in Fig.~\ref{fig:dis}. Obviously,  the circumnavigation radius controlled by our method converges faster than that of the VF guidance law. Fig.~\ref{fig:Fisher} compares the accumulative   information of the two methods before the UAV achieves the desired circumnavigation.   Note that after achieving the circumnavigation with the desired
		radius, the UAVs controlled by the two methods gain  the same
		accumulative information in unit time as illustrated in Fig.~\ref{fig:Q}. Therefore, in order to clearly illustrate the
		difference of the accumulative   information obtained by
		the two methods, Fig.~\ref{fig:Fisher} only illustrates the first 25  seconds of the accumulative  information.    It can be observed from Fig.~\ref{fig:Fisher} that the accumulative   information acquired by our method is higher than that of the VF method. In order to show this more clearly, we demonstrate the obtained accumulative information   per simulation step, i.e., the value of the function $\hat{Q}(r_h,\eta)$, of the two methods   in Fig.~\ref{fig:Q}. Overall, the UAV controlled by our method  gains more accumulative information   before the UAV achieves the desired circumnavigation (except for the first 3 seconds). After the UAV achieves the desired circumnavigation, the values of the function $\hat{Q}(r_h,\eta)$ obtained by the two methods are the same.

	%\begin{figure}[t]
	%	\centering  
	%
	%\end{figure}
	%
	%
	%\begin{figure}[t]
	%	\centering  
	% 
	%\end{figure}
	
	\section{Conclusions} \label{sec:6}
	
 	In this paper, we have addressed the optimal control problem for  NF nonlinear systems with unsymmetrical and state-dependent  input constraint.  The  method  proposed in this paper relaxes the assumptions on the dynamics  and the input constraints of optimal control systems in the existing works.    The proposed method is applied to solve an application case: the optimal UAV circumnavigation control problem. The control performance of our algorithm  has been compared with the algorithm proposed in \cite{dong2019flight} by using a numerical simulation.  
	
	  In the future work, we will   extend the proposed optimal control design method to  the multi-agent systems and further investigate the optimal cooperative circumnavigation control problem of the multi-UAV systems.

	\ifCLASSOPTIONcaptionsoff
	\newpage
	\fi
	
	\bibliographystyle{ieeetr}
	\bibliography{ref/refs}
	
 \begin{IEEEbiography}[{\includegraphics[width=1in,height=1.25in,clip,keepaspectratio]{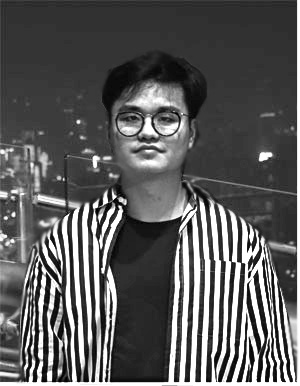}}]{Yangguang Yu}
 	
 	received the B.S., M.S., and Ph.D. degrees from National University of
 	Defense Technology, China, in 2015,  2018, and 2022,
 	respectively. He is currently  a Lecturer at  College of Intelligence Science and Technology, National University of Defense Technology at Changsha. His research interests
 	include multi-agent systems, unmanned aerial vehicles
 	and optimal adaptive control.
 \end{IEEEbiography}
 
 \vspace*{-20\baselineskip}
 % if you will not have a photo at all:
 \begin{IEEEbiography}[{\includegraphics[width=1in,height=1.25in,clip,keepaspectratio]{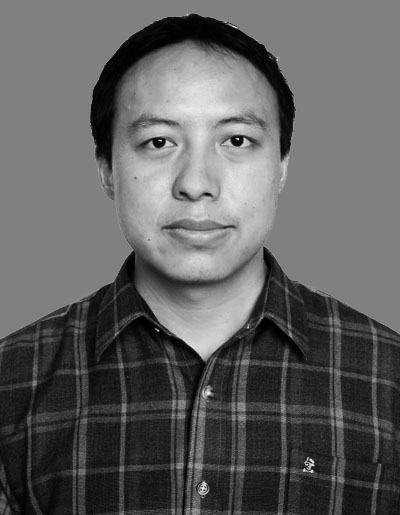}}]{Xiangke Wang}
 	(SM'18) received the B.S., M.S., and Ph.D. degrees in Control Science and Engineering from National University of Defense Technology, China, in 2004, 2006 and 2012, respectively. From 2012, he served as a Lecturer, Associate professor and Professor with the College of Intelligence Science and Technology, National University of Defense Technology, China. He was a visiting student at the Research School of Engineering, Australian National University  from 2009 to 2011. 
 	
 	His current research interests focus on the control of multi-agent systems and its applications on unmanned aerial vehicles. He has authored or coauthored 2 books and more than 100 publications in peer reviewed journals and international conferences, including IEEE Transactions, IJRNC, CDC, IFAC, ICRA. etc.
 \end{IEEEbiography}
 
 \vspace*{-20\baselineskip}
 
 \begin{IEEEbiography}[{\includegraphics[width=1in,height=1.25in,clip,keepaspectratio]{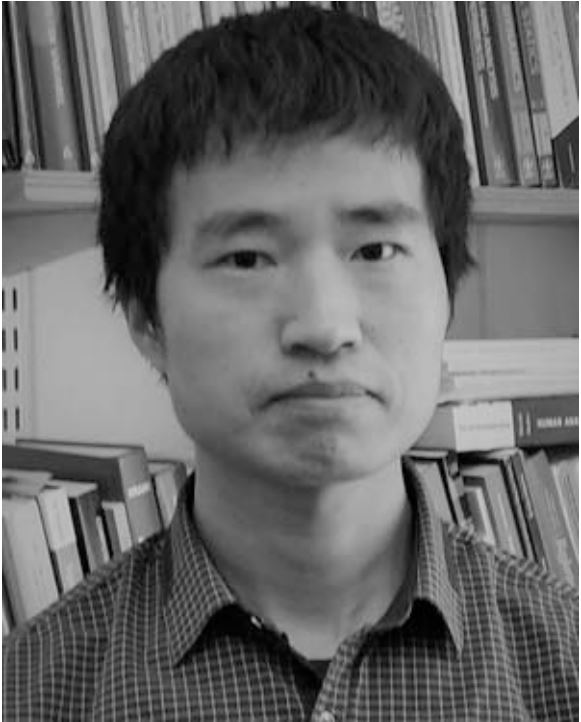}}]{Zhiyong Sun}
 received the Ph.D. degree from The Australian National University (ANU), Canberra ACT, Australia, in February 2017. He was a Research Fellow/Lecturer with the Research School of Engineering, ANU, from 2017 to 2018. From June 2018 to January 2020, he worked as a postdoctoral researcher at Department of Automatic Control, Lund University, Lund, Sweden. Since January 2020 he has joined Eindhoven University of Technology (TU/e), the Netherlands, as an assistant professor. His research interests include multi-robotic systems, control of autonomous formations, distributed control and optimization.
 \end{IEEEbiography}
 \vspace*{-20\baselineskip}
 %\vspace{-140 mm} 
 \begin{IEEEbiography}[{\includegraphics[width=1in,height=1.25in,clip,keepaspectratio]{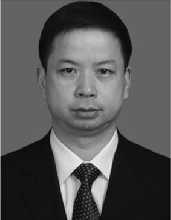}}]{Lincheng Shen}
 	received the B.S., M.S., and Ph.D.
 	degrees in automatic control from the National
 	University of Defense Technology, China, in 1986,
 	1989, and 1994, respectively. In 1989, he joined the
 	Department of Automatic Control, NUDT, where he
 	is currently a full professor and serves as the Dean
 	of the Graduate School. He has been serving as an
 	Editorial Board Member of the Journal of Bionic
 	Engineering since 2007. His research interests include
 	unmanned aerial vehicles, swarm robotics, and
 	artificial intelligence.
 \end{IEEEbiography}

	% that's all folks
\end{document}